\documentclass[11pt,a4paper]{article}
\pdfoutput=1
\usepackage{jheppub,bm,booktabs,multirow}
\usepackage{color}
\usepackage{overpic}
\allowdisplaybreaks

\usepackage[normalem]{ulem}
\usepackage{color}
\usepackage{overpic}

\allowdisplaybreaks

\usepackage[utf8x]{inputenc}

\makeatletter
\def\@fpheader{~}
\makeatother

\def\e{\epsilon}
\def\nno{\nonumber}

\definecolor{military}{rgb}{0.0, 0.5, 0.0}

\preprint{\begin{flushright}
CERN-TH-2019-006
\end{flushright}}

\title{ NLL${'}$ resummation of jet mass}
\author[a]{Marcel Balsiger}
\author[a]{\!, Thomas Becher}
\author[b]{and Ding Yu Shao}

\affiliation[a]{Albert Einstein Center for Fundamental Physics, Institut f\"ur Theoretische Physik, Universit\"at Bern,
  Sidlerstrasse 5, CH-3012 Bern, Switzerland}
  \affiliation[b]{CERN, Theoretical Physics Department, CH-1211, Geneva 23, Switzerland}

\emailAdd{balsiger@itp.unibe.ch}
\emailAdd{becher@itp.unibe.ch}
\emailAdd{dingyu.shao@cern.ch}

\date{\today}

\abstract{Starting from a factorization theorem in effective field theory, we present resummed results for two non-global observables: the invariant-mass distribution of jets and the energy distribution outside jets. Our results include the full next-to-leading-order corrections to the hard, jet and soft functions and are implemented in a parton-shower framework which generates the renormalization-group running in the effective theory. The inclusion of these matching corrections leads to an improved description of the data and reduced theoretical uncertainties. They will have to be combined with two-loop running in the future, but our results are an important first step towards the higher-logarithmic resummation of non-global observables.}

\begin{document}

\maketitle

\section{Introduction}

Up to now, higher-logarithmic resummations of collider observables have only been performed for the narrow class of {\em global} observables which constrain radiation uniformly over the entire phase space. This category includes very inclusive observables such as selected event shapes, but it excludes all observables with hard phase-space cuts or a fixed number of jets. In recent years, a lot of progress was made in the theoretical analysis of non-global observables \cite{Hatta:2008st,Avsar:2009yb,Hatta:2013iba,Caron-Huot:2015bja,Larkoski:2015zka,Becher:2015hka,Neill:2015nya,Becher:2016mmh,Larkoski:2016zzc,Becher:2016omr,Neill:2016stq,Becher:2017nof,Hatta:2017fwr,Martinez:2018ffw,Balsiger:2018ezi,Neill:2018yet}. This includes work on the structure of higher logarithms as well as studies of leading logarithms beyond the large-$N_c$ limit.

In this paper we start the computation of higher-logarithmic terms for non-global observables by analyzing two simple observables, the jet mass and the interjet energy flow, and presenting resummed predictions which include the full one-loop corrections to the relevant hard scattering processes, as well as the associated jet and soft functions. In the effective-theory framework we use for resummation \cite{Becher:2015hka,Becher:2016mmh}, these correspond to matching corrections and they will need to be supplemented by corrections to the renormalization-group (RG) running in the future to arrive at a complete higher-logarithmic treatment of the non-global part. 

Our main goal in the present work is to develop the Monte Carlo methods to include these corrections as a step towards full higher-logarithmic resummation, but it is also interesting to study their numerical size, since they have never been computed for non-global observables and often dominate numerically in the global case. It is customary to add a prime to the logarithmic accuracy to indicate the presence of higher-order matching corrections. In this notation our next-to-leading-logarithmic results for the jet mass have NLL$\rm '$ accuracy.

In Refs.~\cite{Becher:2016mmh,Becher:2016omr} we have derived a factorization formula for interjet energy flow and light-jet mass. The key element is the presence of multi-Wilson-line operators which generate the intricate pattern of Non-Global Logarithms~(NGLs). Explicitly, the result for interjet energy flow at a lepton collider has the form
\begin{align}\label{sigbarefinal}
\sigma(Q, Q_0) &=  \sum_{m=2}^\infty \big\langle \bm{\mathcal{H}}_m(\{\underline{n}\},Q,\mu) \otimes \bm{\mathcal{S}}_m(\{\underline{n}\},Q_0,\mu) \big\rangle \,,
\end{align}
where $Q$ is the center-of-mass energy, and $Q_0=\beta Q$ is the energy scale above which we veto energy in the gap outside the jet cones. For simplicity, we choose the jet axis along the thrust axis. The above factorization formula neglects power corrections from $\mathcal{O}(\beta)$ terms. The hard functions $\bm{\mathcal{H}}_m$ describe hard radiation inside the jet cone, and their characteristic scale is $Q$ since radiation inside the cones is unrestricted. The index $m$ represents the number of hard partons inside the jet, which propagate along the directions $\{\underline{n}\} = \{n_1, n_2, \dots, n_m\}$. Each of these sources soft radiation, which we describe by a Wilson line along the direction of the hard parton. The matrix elements of these Wilson lines define the soft functions $\bm{\mathcal{S}}_m(\{\underline{n}\},Q_0,\mu)$. To obtain the cross section, one integrates over the directions $\{\underline{n}\}$, which is indicated by the symbol $\otimes$. The hard and soft functions are matrices in the color space of the $m$ partons and one takes the color trace $\langle \dots \rangle$ after multiplying them. The operator definition for these functions and further explanations can be found in \cite{Becher:2016mmh}. 

\begin{figure}[t]
\center
\includegraphics[width=.34\textwidth]{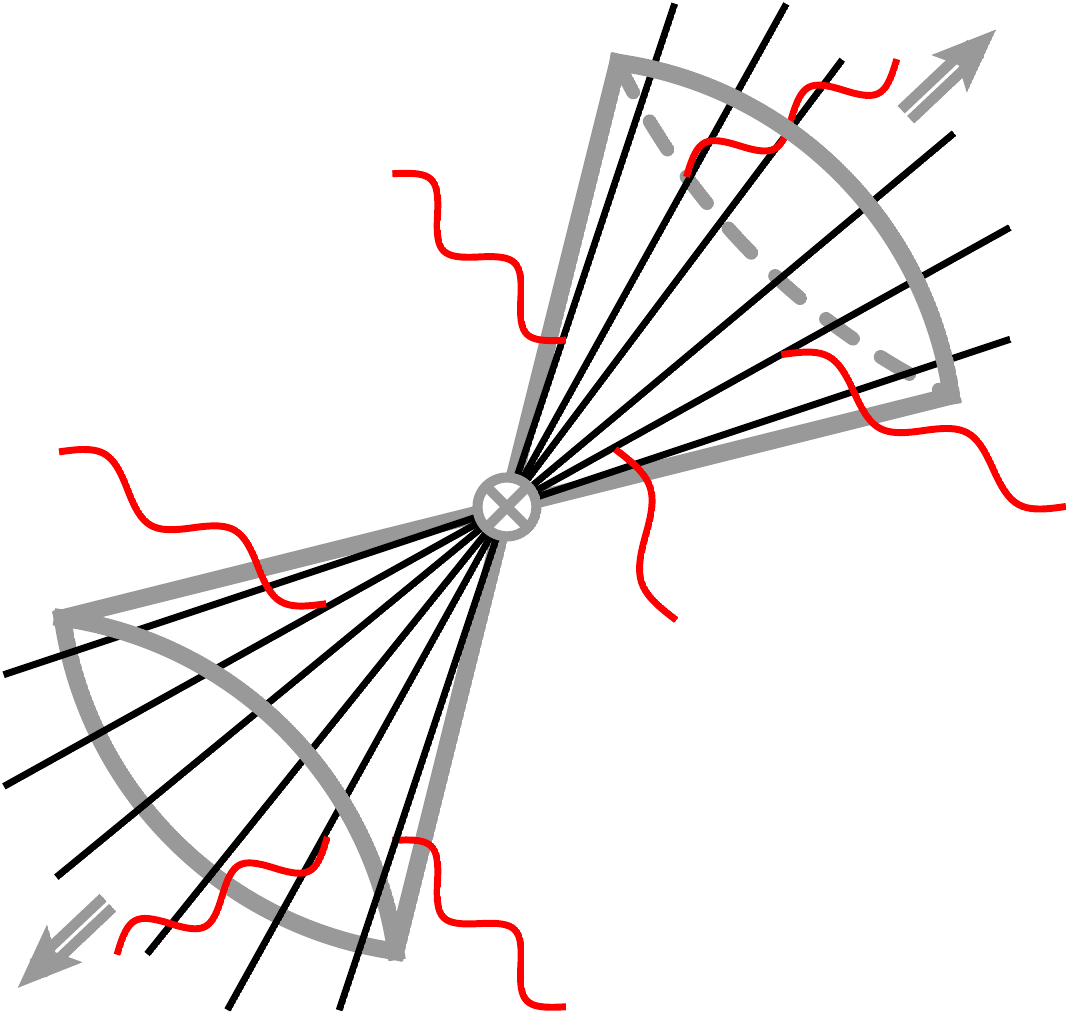}\hspace{1.5cm}
\includegraphics[width=.42\textwidth]{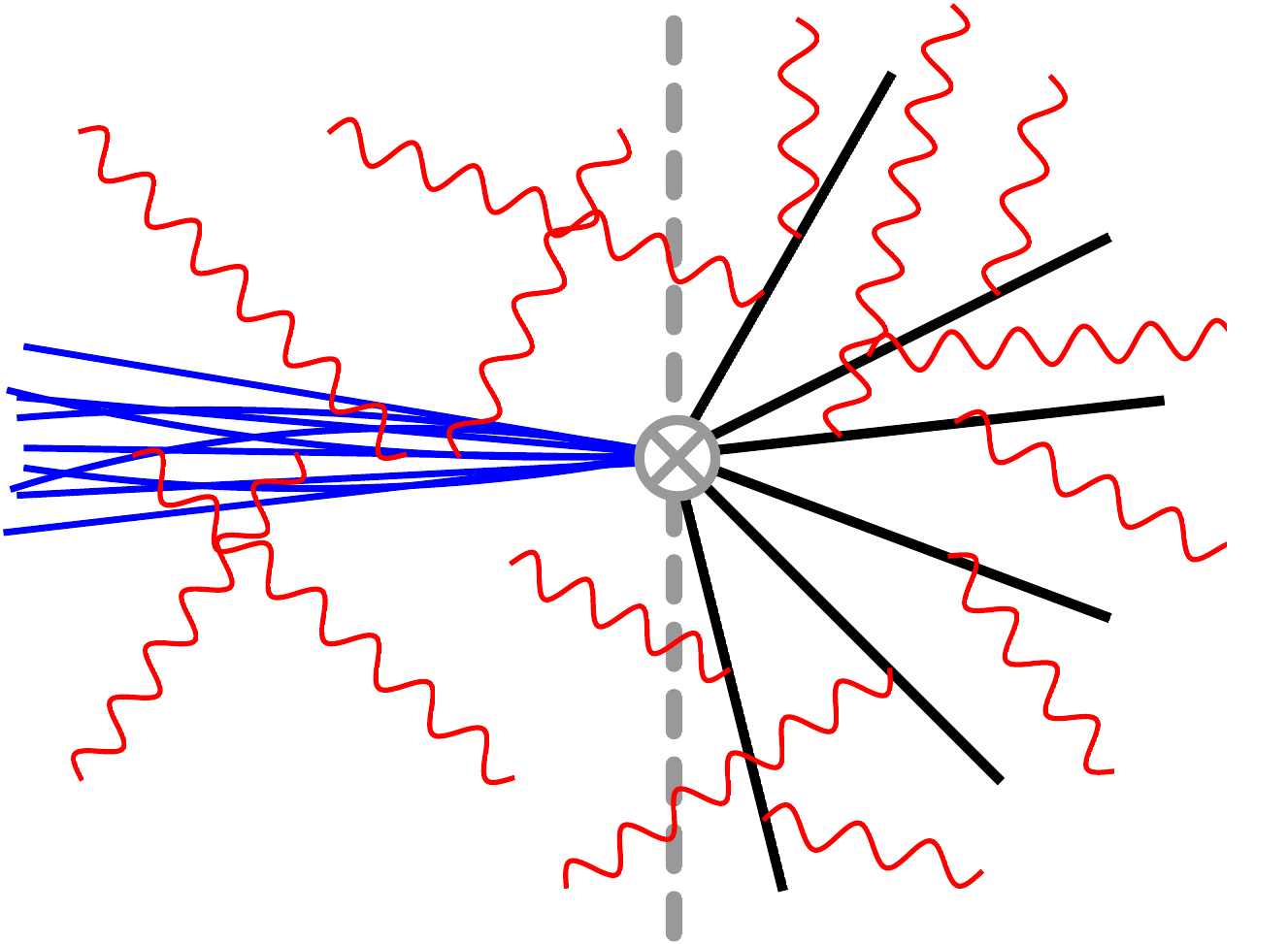}
\caption{Pictorial representations of factorization formulas for interjet energy flow (left) and jet mass (right), see \eqref{sigbarefinal} and \eqref{eq:factlightmassLap}.  The black lines represent hard radiation with typical scale $Q$ which is constrained to be inside the cones, and the red lines depict soft radiation with a low energy scale $Q_0$ which is allowed to populate the full phase space. In the right figure, the blue lines in the left hemisphere represent collinear radiation which is described by the inclusive jet function in \eqref{eq:factlightmassLap}. \label{fig:intro}}
\end{figure}

The second observable we consider is the jet mass distribution at a lepton collider. To define the jet mass, we use the thrust axis to split every event into two hemispheres. One can then (randomly) select one of the two jets and compute its invariant mass $M$, which is usually discussed in terms of the dimensionless variable $\rho = M^2/Q^2$.  Alternatively, one computes the mass in both hemispheres and chooses the heavier mass $\rho_h$ or lighter one $\rho_\ell$. Obviously, there is a relation among the these observables: the jet mass distribution is simply the average of heavy-jet mass and light-jet mass one
\begin{align}\label{eq:jetmassrel}
\frac{d\sigma}{d\rho} = \frac{1}{2} \left( \frac{d\sigma}{d\rho_\ell} + \frac{d\sigma}{d\rho_h} \right).
\end{align}
We will call the hemisphere we select to measure the mass the \textit{left} one, which means that the radiation in the \textit{right} hemisphere is unconstrained.\footnote{In our previous paper, we called $\rho$ the left-jet mass and denoted it by $\rho_L$ \cite{Becher:2016omr}.} We introduce a light-like reference four-vector $n^\mu=(1,0,0,1)$ pointing to the right along the thrust axis and an opposite vector $\bar n^\mu=(1,0,0,-1)$ pointing to the left. The hard partons in the right hemisphere then generate the complicated pattern of soft radiation and associated NGLs. The main difference to formula (\ref{sigbarefinal}) is that one also needs the standard inclusive jet functions to describe collinear radiation in the left hemisphere. Resummation effects in the jet mass distribution have been discussed in Refs.~\cite{Dasgupta:2012hg,Chien:2012ur,Liu:2014oog,Idilbi:2016hoa,Kang:2018jwa}, however only in \cite{Dasgupta:2012hg} the leading NGLs were resummed. Our work is based on the factorization theorem for jet mass derived in \cite{Becher:2016omr}.  The invariant mass of the left jet is obtained from the momentum $p_{\bar c}$ of the energetic particles collinear to $\bar{n}$ and the soft partons in the left hemisphere,
\begin{equation}
\rho\, Q^2 =M^2 = (p_{\bar c} + p_s)^2 = p_{\bar c}^2 + Q\, \bar{n}\cdot p_s +\mathcal{O}(p_s^2)\,.
\end{equation}
In the factorization theorem, the sum results in a convolution of the soft and jet functions. To avoid this, one can work in Laplace space, where the factorization formula has the product form
\begin{equation}\label{eq:factlightmassLap}
\tilde\sigma(\tau)  = \sum_{i=q,\bar{q},g} \tilde{j}_i(\tau Q,\mu) \, \sum_{m=1}^\infty \big\langle \bm{\mathcal{H}}^i_m(\{\underline{n}\}, Q,\mu) \otimes \bm{\mathcal{\widetilde{S}}}_m(\{\underline{n}\},\tau,\mu) \big\rangle\,,
\end{equation}
where $\tau$ is the Laplace conjugate variable of $\rho$, and $\tilde j_i$ is the inclusive jet function \cite{Becher:2006qw,Becher:2010pd}, which by now is known to three loops~\cite{Bruser:2018rad, Banerjee:2018ozf}. In \eqref{eq:factlightmassLap}  the index $m$ indicates the number of partons in the  inclusive (right) hemisphere, so that $m=1$ at leading order (LO).

As long as we consider large jet cone sizes of $\mathcal{O}(1)$, the leading-logarithms (LLs) in interjet energy flow at a lepton collider are of the form $\alpha_s^n \ln^n\beta$. The  interjet energy flow is a single logarithmic observable, because collinear logarithms cancel inside the large cone region and only soft logarithms remain. These logarithms arise from the multi-Wilson-line operators $\bm{\mathcal{S}}_m$ in \eqref{sigbarefinal} and one needs to use parton shower methods to resum the enhanced logarithms already at the LL level.  In \cite{Balsiger:2018ezi} we have written a dedicated parton-shower code to perform the resummation for such observables and have interfaced it with the {\sc MadGraph5$\_$aMC@NLO} event generator \cite{Alwall:2014hca}. This provides an automated framework to perform the LL resummation for single-logarithmic observables. However, collider observables are typically double logarithmic. The leading logarithms in the jet mass distribution, for example, are $\alpha_s^n \ln^{2n}\rho$. Even for non-global observables, these double logarithmic terms have a simple structure, and they can be factored out and treated separately. In the parton shower framework, we therefore subtract these ``global'' contributions and exponentiate them manually, as Dasgupta and Salam did in their original paper on NGLs \cite{Dasgupta:2001sh}. Given their different nature, it is interesting to analyze both the interjet energy flow and the jet mass as examples and we will present LL$\rm '$ and NLL$\rm'$ improved results for single logarithmic and double logarithmic observables, separately. A second motivation to also analyze the jet mass, is that there are {\sc LEP} measurements to which we can compare to, in contrast to the interjet energy flow. Unfortunately, the typical jet mass at LEP jet is quite low  $M\lesssim 10\, {\rm GeV}$, which translates to a scale of the soft radiation of $Q_0 \sim M^2/Q  \lesssim 1\, {\rm GeV}$ so that non-perturbative effects are very important in the peak region of the distribution.

Our paper is organized as follows. In the next section, we will discuss ${\rm LL'}$ resummation for interjet energy flow and show how one implements the one-loop corrections to the hard and soft functions. We then move to the jet mass distribution in Section \ref{sec:jetmass}, focussing on the differences to the single-logarithmic case. We will in particular show how to subtract global logarithms in the parton shower and in the soft function. After presenting numerical results in Section~\ref{sec:numerics} and comparing to {\sc LEP} data and {\sc PYTHIA} results, we conclude in Section \ref{sec:concl}.

\section{Interjet energy flow at ${\rm LL}'$ accuracy\label{sec:interjet}}

The perturbative expansion of the  interjet energy flow in \eqref{sigbarefinal} suffers from large logarithms of the ratio of the hard scale $Q$ and the soft scale $Q_0$. To resum these, one solves the RG equation of the hard function and evolves it from its characteristic scale $\mu_h \sim Q$ down to a soft scale  $\mu_s \sim Q_0$. This yields the RG-improved expression \cite{Becher:2016mmh}
\begin{align}\label{eq:sigmaRGimproved}
  \sigma(Q, Q_0) 
   = \sum_{l=2}^\infty \big\langle \bm{\mathcal{H}}_l(\{\underline{n}'\},Q,\mu_h) 
    \otimes \sum_{m\geq l}^\infty \bm{U}_{lm}(\{\underline{n}\},\mu_s,\mu_h)\,\hat{\otimes}\, 
    \bm{\mathcal{S}}_m(\{\underline{n}\},Q_0,\mu_s) \big\rangle \,,
\end{align}
where the evolution factor is defined as a path-ordered exponential of the anomalous dimension
\begin{align}
\bm{U}(\{\underline{n}\},\mu_s,\mu_h) = {\rm \bf P} \exp\left[ \int_{\mu_s}^{\mu_h} \frac{d\mu}{\mu} \bm{\Gamma}^{H}( \{\underline{n}\} ,\mu) \right].
\end{align}
The RG-evolution generates additional partons and maps the $l$-parton configuration along the directions $\{\underline{n}'\}= \{n_1 , \dots , n_l\} $ into an $m$-parton final state along the directions $\{\underline{n}\}=\{n_1 , \dots , n_l, n_{l+1}, \dots, n_m\}$. The symbol $\hat{\otimes}$ in \eqref{eq:sigmaRGimproved} indicates the integral over the directions of the additional $m-l$ partons generated in the evolution.

At the leading logarithmic level, we only need the one-loop anomalous dimension and can rewrite the exponent as
\begin{equation}
\int_{\mu_s}^{\mu_h} \frac{d\mu}{\mu}\, \bm{\Gamma}^H= \int_{\alpha_s(\mu_s)}^{\alpha_s(\mu_h)} \frac{d\alpha}{\beta(\alpha)}\, \frac{\alpha}{4\pi}\,\bm{\Gamma}^{(1)}=  \frac{1}{2\beta_0}\ln\frac{\alpha_s(\mu_s)}{\alpha_s(\mu_h)} \,\bm{\Gamma}^{(1)}  \equiv t \,\bm{\Gamma}^{(1)}  \,.
\end{equation}
In the last step, we have introduced the evolution time $t\equiv t(\mu_h,\mu_s)$. For a given $\mu_h$, there is a one-to-one correspondence of the evolution time to the low scale $\mu_s$. Obviously, for $\mu_h=\mu_s$, we have $t=0$. During the evolution, $t$ grows and goes to infinity as $\mu_s$ hits the Landau pole. For $\mu_h = M_Z$ and two-loop running with a Landau pole at $\Lambda = 0.230\,{\rm GeV}$, the choice $\mu_s = 1\, {\rm GeV}$ corresponds to $t=0.08$. A plot connecting $t$ and $\mu_s$ for different values of $\mu_h$ can be found in Figure 1 of our previous paper \cite{Balsiger:2018ezi}.

In \cite{Balsiger:2018ezi} we implemented the RG evolution factor $ \bm{U}(\{\underline{n}\},\mu_s,\mu_h) 
$ in the large-$N_c$ limit using the parton shower method proposed by Dasgupta and Salam in \cite{Dasgupta:2001sh}.  We don't want to repeat the entire discussion here, but we give the algorithm in Appendix~\ref{IJmc}, since we need to extend it to compute the soft functions, as discussed below. Let us also list the one-loop anomalous dimension, since its form will be relevant in the discussion of the jet mass below. It is given by \cite{Becher:2016mmh}
\begin{equation}\label{eq:gammaOne}
\bm{\Gamma}^{(1)} =  \left(
\begin{array}{ccccc}
   \, \bm{V}_{2} &   \bm{R}_{2} &  0 & 0 & \hdots \\
 0 & \bm{V}_{3} & \bm{R}_{3}  & 0 & \hdots \\
0 &0  &  \bm{V}_{4} &  \bm{R}_{4} &   \hdots \\
 0& 0& 0 &  \bm{V}_{5} & \hdots
   \\
 \vdots & \vdots & \vdots & \vdots &
   \ddots \\
\end{array}
\right).
\end{equation}
The entries $\bm{R}_m$ and $\bm{V}_m$ are angular functions associated with the emission of a real or virtual soft gluon and take the form
\begin{align}\label{eq:oneLoopRG}
 \bm{V}_m  &= 2\,\sum_{(ij)}\,(\bm{T}_{i,L}\cdot  \bm{T}_{j,L}+\bm{T}_{i,R}\cdot  \bm{T}_{j,R})  \int \frac{d\Omega(n_k)}{4\pi}\, W_{ij}^k,  \nonumber\\
 \bm{R}_m & = -4\,\sum_{(ij)}\,\bm{T}_{i,L}\cdot\bm{T}_{j,R}  \,W_{ij}^{m+1}\,  \Theta_{\rm in}(n_{m+1})\,,  
\end{align}
where the color matrices $\bm{T}_{i,L}$ act on the hard function from the left, i.e. on the amplitude, while $\bm{T}_{i,R}$ acts on the conjugate amplitude. The sum runs over all unequal pairs $(ij)$ of the $m$ hard partons. The anomalous dimension involves the dipole radiator
\begin{equation}\label{eq:dipole}
W_{ij}^k = \frac{n_i \cdot n_j}{(n_i \cdot n_k) (n_j \cdot n_k)}\,,
\end{equation}
which is given by the product of the associated eikonal factors. In the virtual corrections, one integrates over the direction $n_k$ of the emission. We note that individually $\bm{R}_m$ and $\bm{V}_m$ suffer from collinear divergences, which cancel in the cross section. In the Monte Carlo implementation, one works with a collinear cutoff to regularize the divergences.

As long as we choose the $\mu_h$ and $\mu_s$ properly, the hard and soft functions will be free of large logarithms and the large logarithmic terms are resummed in the evolution factor. Because they are free of large logarithms, the higher-multiplicity hard functions are suppressed by $\alpha_s$ as $\bm{\mathcal{H}}_l \sim \alpha_s^{l-2} \bm{\mathcal{H}}_2$. At LL level, we thus only need to include the hard function $\bm{\mathcal{H}}_2$ and the soft function is given as the unit matrix in the color space $\bm{\mathcal{S}}_m \sim \bm{1}$. At  LL accuracy, the RG-improved result  \eqref{eq:sigmaRGimproved} simplifies to
\begin{align}\label{eq:LLIJ}
  \sigma^{\rm LL}(Q, Q_0) 
   = \sum_{m = 2}^\infty \big\langle \bm{\mathcal{H}}_2(\{n_1,n_2\},Q,\mu_h) 
    \otimes  \bm{U}_{2m}(\{\underline{n}\},\mu_s,\mu_h)\,\hat{\otimes}\, 
    \bm{1} \big\rangle \,.
\end{align}

\begin{figure}[t]
\hspace{1.6cm}
\begin{overpic}[scale=0.4]{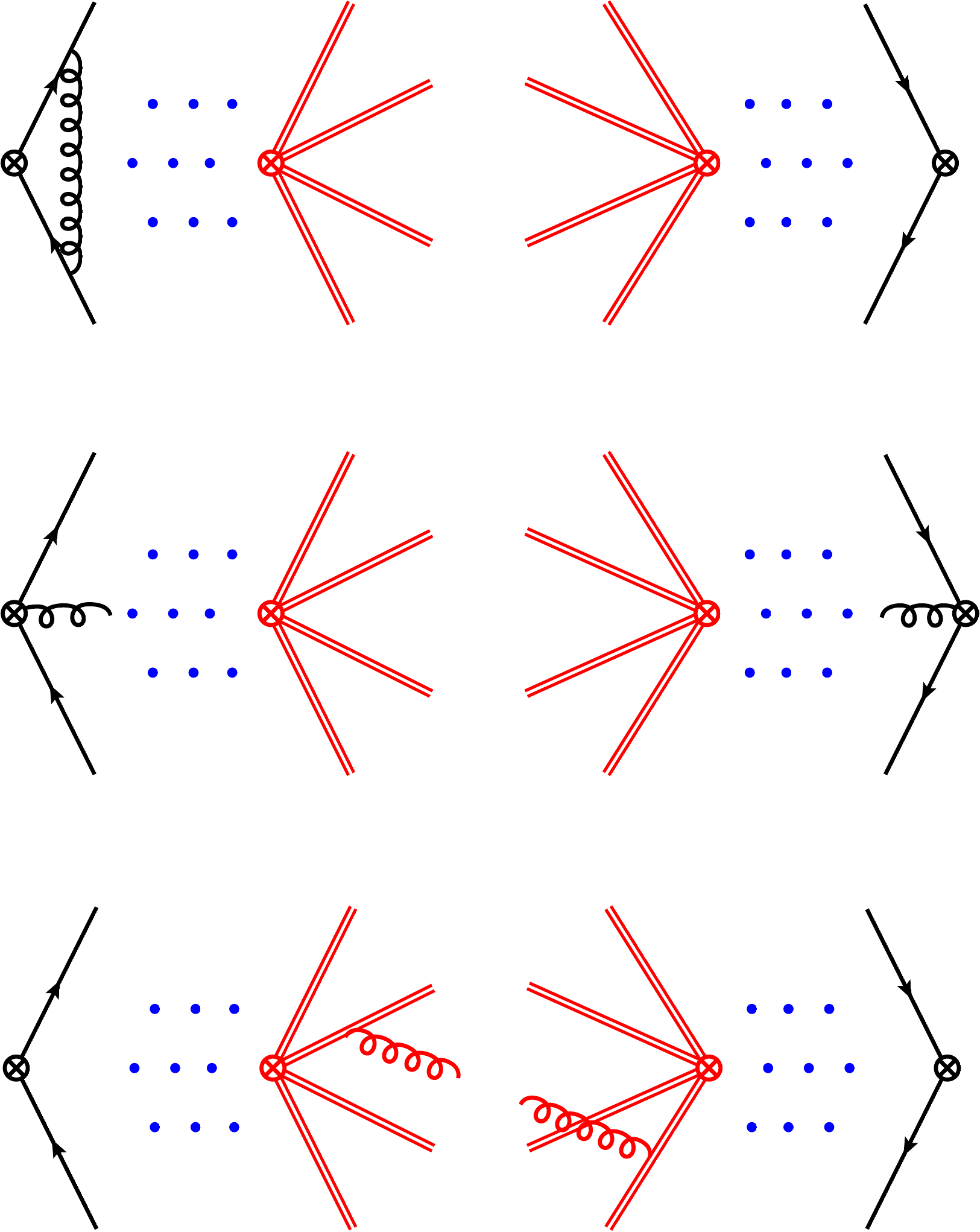}
\put(85,86){ $\large \sim ~~~~  \bm{\mathcal{H}}_2^{(1)} \, \otimes \,  {\color{blue} \bm{U}_{2m} }\,\hat{\otimes} \, \color{red}{\bm{\mathcal{S}}_m^{(0)} } $}
\put(85,50){ $\large \sim ~~~~ \bm{\mathcal{H}}_3^{(1)} \, \otimes \,  {\color{blue} \bm{U}_{3m} }\,\hat{\otimes} \, \color{red}{\bm{\mathcal{S}}_m^{(0)} } $}
\put(85,12){ $\large \sim ~~~~ \bm{\mathcal{H}}_2^{(0)} \, \otimes \,  {\color{blue} \bm{U}_{2m} }\,\hat{\otimes} \, \color{red}{\bm{\mathcal{S}}_m^{(1)} } $}
\end{overpic}
\caption{Pictorial representations of the different ingredients for $\rm LL'$ resummation of the interjet energy flow. The diagrams on the three lines correspond to the one-loop corrections from  $\bm{\mathcal{H}}_2^{(1)}$, $\bm{\mathcal{H}}_3^{(1)}$ and $\bm{\mathcal{S}}_m^{(1)}$, respectively. The virtual corrections to $\bm{\mathcal{S}}_m$ are scaleless and vanish. \label{fig:InterJ_LLp}}
\end{figure}

To extend these results to NLL, one needs two ingredients: the one-loop matching corrections and the corrections to the RG running due to the two-loop anomalous dimensions. The present paper focuses on the first set of corrections, i.e.  ${\rm LL'}$ accuracy. Specifically, we need one-loop corrections to $\bm{\mathcal{H}}_2$, the tree-level result for $\bm{\mathcal{H}}_3$ and the one-loop soft functions $\bm{\mathcal{S}}_m$.  We write their perturbative expansions in the form
\begin{align}
\bm{\mathcal{H}}_2 & = \sigma_0 \left(\bm{\mathcal{H}}_2^{(0)} + \frac{\alpha_s}{4\pi}\bm{\mathcal{H}}_2^{(1)} + \cdots \right), ~~~~~~
\bm{\mathcal{H}}_3  = \sigma_0  \left(\frac{\alpha_s}{4\pi}\bm{\mathcal{H}}_3^{(1)} + \cdots \right), \nno \\
\bm{\mathcal{S}}_m & = \bm{1} + \frac{\alpha_s}{4\pi} \bm{\mathcal{S}}_m^{(1)} + \cdots .
\end{align}
In this notation, the full ${\rm LL'}$ resummed cross section takes the form
\begin{equation}
\begin{aligned}\label{LLp}
\frac{\sigma^{\rm LL'}(Q,Q_0)}{\sigma_0} =\, & \sum_{m=2}^\infty \big\langle  \bm{\mathcal{H}}_2^{(0)}(\{n_1,n_2 \},Q,\mu_h) \, \otimes \,  \bm{U}_{2m}(\{\underline{n}\},\mu_s,\mu_h) \,\hat{\otimes}\, \bm{1} \big\rangle   \\
 &\hspace{-1.5cm}+ \frac{\alpha_s(\mu_h)}{4\pi}  \sum_{m=2}^\infty \big\langle  \bm{\mathcal{H}}_2^{(1)}(\{n_1,n_2\},Q,\mu_h)  \, \otimes \,  \bm{U}_{2m}(\{\underline{n}\},\mu_s,\mu_h) \,\hat{\otimes}\, \bm{1} \big\rangle   \\
&\hspace{-1.5cm} + \frac{\alpha_s (\mu_h)}{4\pi} \sum_{m=3}^\infty \big\langle  \bm{\mathcal{H}}_3^{(1)}(\{n_1,n_2,n_3\},Q,\mu_h)  \, \otimes \,  \bm{U}_{3m}(\{\underline{n}\},\mu_s,\mu_h) \,\hat{\otimes}\, \bm{1} \big\rangle   \\
&\hspace{-1.5cm} + \frac{\alpha_s (\mu_s)}{4\pi}\sum_{m=2}^\infty \big\langle   \bm{\mathcal{H}}_2^{(0)}(\{n_1,n_2\},Q,\mu_h)  \, \otimes \,  \bm{U}_{2m}(\{\underline{n}\},\mu_s,\mu_h) \,\hat{\otimes}\, \bm{\mathcal{S}}_m^{(1)} (\{\underline{n} \},Q_0,\mu_s)   \big\rangle. 
\end{aligned}
\end{equation}
We used here that the leading-order soft function $ \bm{\mathcal{S}}_m^{(0)}$ is the unit matrix $\bm{1}$ in color space. The first line contains the LL result \eqref{eq:LLIJ}, and the remaining three lines show the different NLO corrections, which are depicted in Figure~\ref{fig:InterJ_LLp}.

The hard functions $\bm{\mathcal{H}}_m$ include the momentum conservation and phase-space constraints on the hard partons. For two partons, these constraints render the integrals over the parton directions trivial. The momentum and jet direction constraints impose that the vectors $n_1$ and $n_2$ must point along the thrust axis and in opposite directions so that
\begin{equation}
\langle \bm{\mathcal{H}}_2 (\{n_1,n_2 \},Q,\mu) \otimes \bm{\mathcal{S}}_2 (\{n_1,n_2 \},Q_0,\mu)\rangle = \sigma_0 \,H_2(Q^2,\mu) \langle \bm{\mathcal{S}}_2 (\{\bar{n}, n \},Q_0,\mu)\rangle \,,
\end{equation}
where we have used that also the color structure is trivial for two hard partons. The function $H_2(Q^2,\mu)$ is the standard dijet hard function
\begin{equation}
H_2(Q^2,\mu) = 1+\frac{\alpha_s}{4\pi} C_F\left[ - 8 \ln^2 \frac{\mu}{Q} - 12 \ln\frac{\mu}{Q} - 16 + \frac{7}{3} \pi^2 \right]\,,
\end{equation}
which arises also for global observables such as the event shape thrust. In the large-$N_c$ limit, we should replace $C_F \to N_c/2$. 

\begin{figure}[t!]
\begin{center}
\hspace*{0.6cm}
\begin{overpic}[scale=0.85]{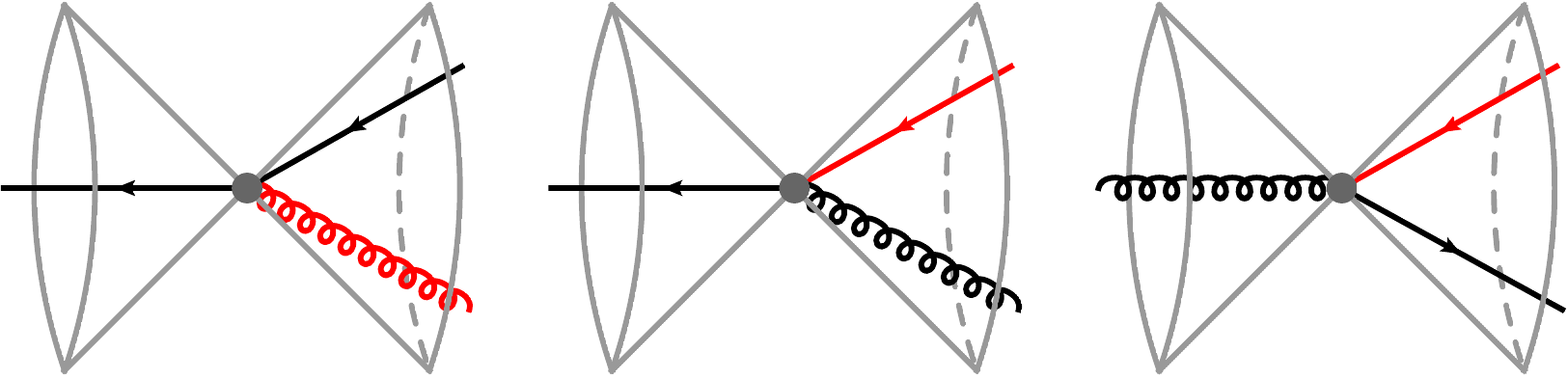}
\put(7,-3){I: $E_1>E_2>E_3$}
\put(40.5,-3){II : $E_1>E_3>E_2$}
\put(75,-3){III:  $E_3>E_1>E_2$}
\put(-6,14){$ q (p_1)$}
\put(29.5,22){$\bar q (p_2)$}
\put(29.5,2){\color{red}{$g(p_3)$}}
\end{overpic}
\end{center}
\vspace{2ex}
\caption{Kinematical configurations in the three different regions with different energy ordering. Particles with the smallest energy are drawn in red.}
\label{fig:H3_dia}
\end{figure}
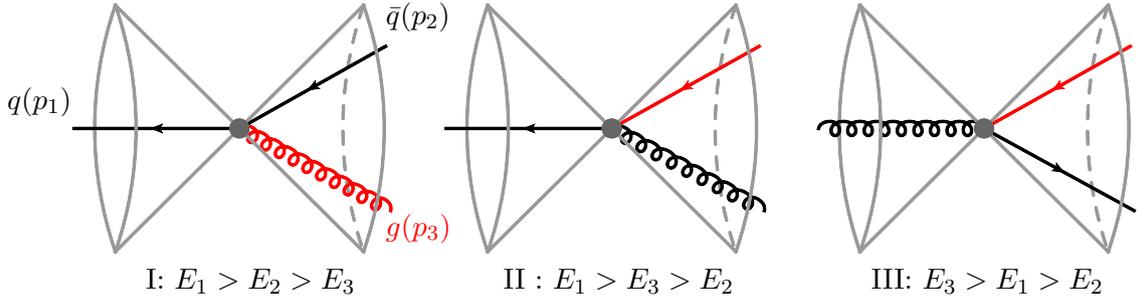

In \cite{Becher:2016mmh} we have derived an expression for the hard function $\bm{\mathcal{H}}_3^{(1)}$, which  corresponds to the QCD process $\gamma^\ast \to q(p_1) \bar q(p_2)  g(p_3)$. By definition $\bm{\mathcal{H}}_3^{(1)}$ only depends on angular information of the three partons, since their energies have already been integrated over. For convenience we split the phase space integration into different regions according to the direction of the thrust axis, which for three-parton final states points in the opposite direction of the most energetic parton. Due to momentum conservation, the three partons must be in a plane. Using invariance of the cross section under rotation around the thrust axis, in Region~I only the angles $\theta_2$ and $\theta_3$, between the partons and the thrust axis, are not fixed. 

For convenience we parameterize these angles in terms of two variables $u$ and $v$ each going from $0$ to $1$ and defined as
%\begin{align}\label{H3Iuv}
%\sin^2\frac{\theta_2}{2} = \frac{\delta^2 u^2 v}{1+\delta^2}, ~~~~~ \sin^2\frac{\theta_3}{2} =\frac{\delta^2 v}{1+\delta^2},
%\end{align}
\begin{align}\label{H3Iuv}
\hat \theta_2 \equiv  \tan \frac{\theta_2}{2}  = u \, v, ~~~~~\hat \theta_3 \equiv \tan\frac{\theta_3}{2} = v,
\end{align}
where the variable $v$ is directly related to the larger angle $\theta_3$, while $u$ characterises the relative size of the angles. Please note that the variables $u$ and $v$ differ from the quantities of the same name used in \cite{Becher:2016mmh}, where we defined the variables such that $v=1$ corresponded to the angle of the jet cone, rather than a $90^\circ$ angle as in \eqref{H3Iuv}. Because the same hard function $\bm{\mathcal{H}}_3^{(1)}$ also arises for the jet mass studied below, we prefer to not incorporate the specific phase-space constraint into its parameterization. 

\begin{figure}[t]
\centering
\begin{tabular}{cc}
\raisebox{0.2cm}{\includegraphics[width=0.46\textwidth]{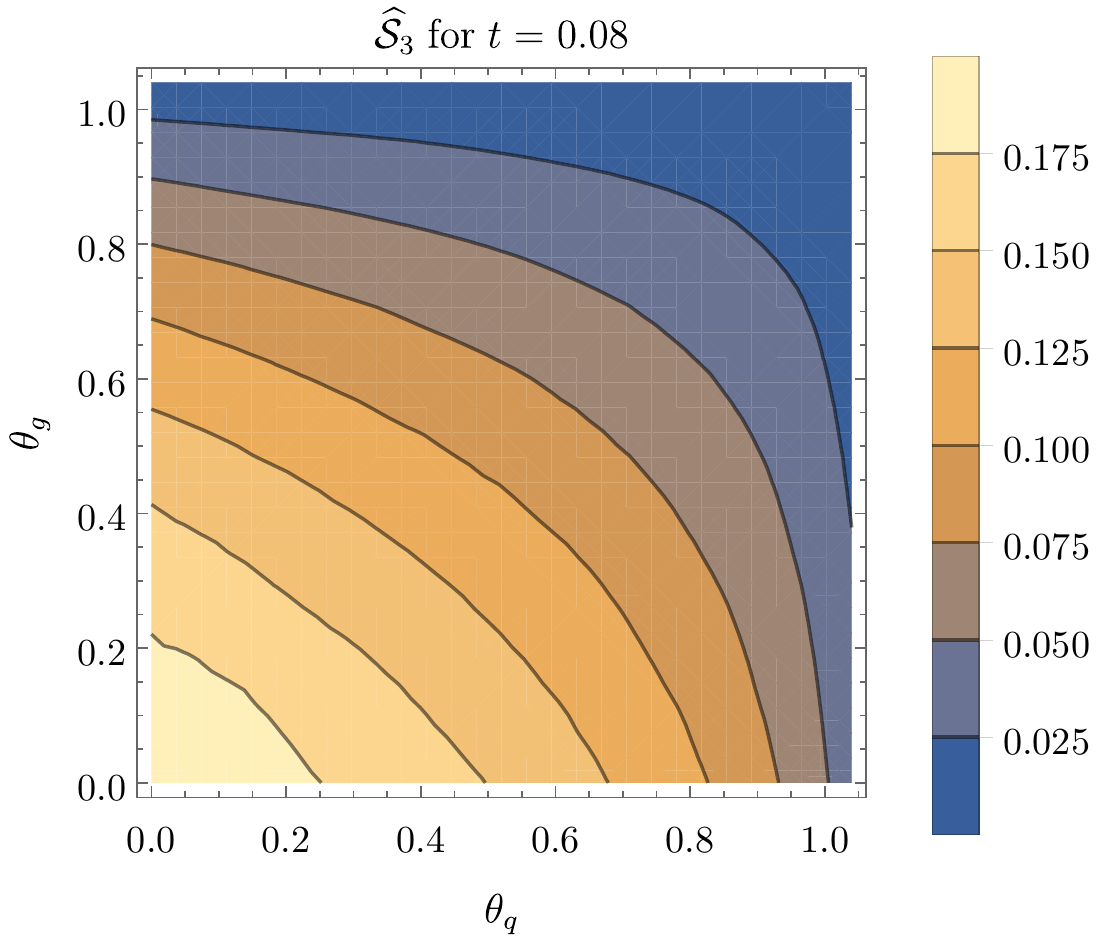}} ~~~~&\raisebox{0.14cm}{ \includegraphics[width=0.413\textwidth]{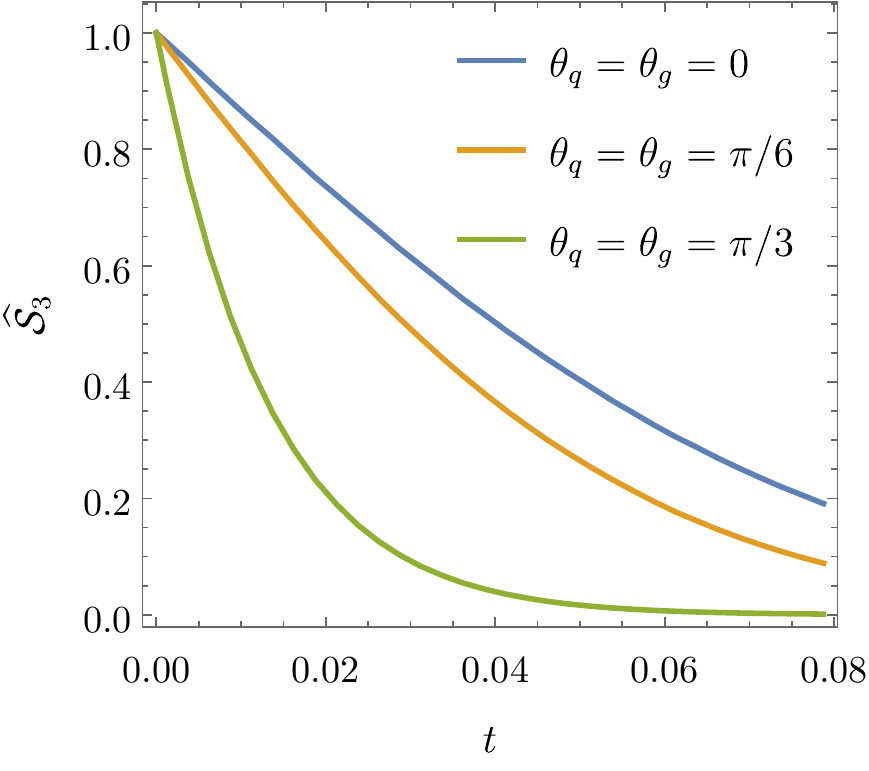}}
 \end{tabular}
 \caption{ Left: Angular dependence of $\widehat{\bm{\mathcal{S}}}_3$ for fixed evolution time $t=0.08$. Note that the angles $\theta_q$ and $\theta_g$  of the hard partons to the jet axis must be smaller than the cone angle $\alpha=\pi/3 \approx 1.04$. Right: Dependence on the evolution time $t$ at fixed angles. \label{fig:S3fun}}
 \end{figure}

The bare hard function $\bm{\mathcal{H}}_3^{(1)}$ in terms of the angles  $\hat \theta_2$ and $\hat \theta_3$ was given in (4.4) of \cite{Becher:2016omr}. The corresponding representation includes a $\theta$-function constraint imposed to prevent the thrust axis from flipping. For simplicity, we choose the jet opening half-angle $\alpha \leq \frac{\pi}{3}$ so that the axis constraint is automatically fulfilled. The hard function suffers from divergences when $u$ and $v$ go to zero. In dimensional regularization after performing $\overline{ \rm MS}$ subtraction, the contribution of Region~I to the renormalized hard function $\bm{\mathcal{H}}_3^{(1)}$ is given by
\begin{align}\label{H3I}
\mathcal{H}_{3, {\rm I}}^{(1)}(u,v, Q,\mu) =&~ C_F \Bigg\{ \left[ 4\ln^2\frac{\mu}{Q} - \frac{\pi^2}{6}\right]\delta(u)\delta(v)  - 8 \ln \frac{\mu}{Q}\delta(u) \left(\frac{1}{v}\right)_+  \nno \\
& \hspace{-2.5cm} + 8\, \delta(u) \left(\frac{\ln v}{v}\right)_+  +\left[ - \ln \frac{\mu}{Q} F(u,0) + \frac{2 u^2}{(1+u)^3} - F(u,0) \ln(1+u) \right] \delta(v) \left(\frac{1}{u}\right)_+  \nno\\
& \hspace{-2.5cm}+ F(u,0) \delta(v) \left(\frac{\ln u}{u}\right)_+ + F(u,v) \left(\frac{1}{u}\right)_+ \left(\frac{1}{v}\right)_+\Bigg\}\Theta_{\rm in}(v).
\end{align}
The function $\Theta_{\rm in}(v)$ ensures that all hard emissions are inside the jet. For the interjet energy flow it is given by $\Theta_{\rm in}(v)=\theta(\delta-v)$, with $\delta=\tan\frac{\alpha}{2}$, where $\alpha$ is the jet opening half-angle. In the large-$N_c$ limit, the color structure of the hard functions becomes trivial and we use non-bold symbols such as $\mathcal{H}_{3, {\rm I}}^{(1)}$ to indicate the scalar quantities which are relevant in this limit. The expression for the auxiliary function $F(u,v)$ is given by 
\begin{align}\label{eq:Ffun}
F(u,v) = \frac{4 \left[u \left[-2 \left(u^2+u+1\right) v^2+u \,(2\, u\, (u+1)+1)\, v^4+u+2\right]+2\right]}{(u+1)^3}  . 
\end{align}
Similarly, in Region~II we have 
\begin{align}\label{H3II}
\mathcal{H}_{3,{\rm II}}^{(1)} (u,v,Q,\mu) =&~ C_F  \Bigg\{ \left[ - \ln \frac{\mu}{Q} G(u,0) +  \frac{2}{(1+u)^3 } +  G(u,0) \ln \left( \frac{u}{1+u} \right)\right]\delta(v) \nno \\
&  \hspace{1.2cm} + G(u,v) \left( \frac{1}{v} \right)_+\Bigg\}\Theta_{\rm in}(v),
\end{align}
%\begin{align}
%\mathcal{H}_{3,{\rm II}}^{(1)} =&~ \frac{N_c}{2}  \Bigg\{ \left[ - 2\ln \frac{\mu}{Q} G(u,0) +  \frac{2}{(1+u)^3 } + G(u,0)  \ln\left(\frac{\delta^2}{1+\delta^2}\right)  + 2 G(u,0) \ln \left( \frac{u}{1+u} \right)\right]\delta(v) \nno \\
%&  \hspace{1.2cm} + G(u,v) \left( \frac{1}{v} \right)_+ \Bigg\},
%\end{align}
with the parametrization $\hat \theta_3 = u v$ and $\hat \theta_2 = v$. The function $G(u,v)$ is defined as
\begin{align}
G(u,v) = \frac{4 \left[u \left[-2 \left(u^2+u+1\right) v^2+u (u (u+2)+2) v^4+2 (u+1)\right]+1\right]}{(u+1)^3} . 
\end{align}
Region~III describes the situation, where the gluon is the most energetic particle and we parameterize $\hat \theta_1 = u v$, $\hat \theta_2 = v$. The hard function reads 
\begin{align}\label{H3III}
\mathcal{H}_{3,{\rm III}}^{(1)}(u,v,Q,\mu) =&~ C_F H(u,v)\Theta_{\rm in}(v),
\end{align}
with 
\begin{align}
H(u,v) =\frac{4 v \left(u^4 v^4+u^2 v^4+4 u^2 v^2+u^2+1\right)}{(u+1)^2
   \left(1-u v^2\right)}. 
\end{align}
Next, we will discuss how to implement the above expressions into the parton shower code. We first rewrite the angular integral in the $\bm{\mathcal{H}}_3^{(1)}$ contribution as
\begin{align}\label{H3S3}
 \big\langle  \bm{\mathcal{H}}_3^{(1)}(\{\underline{n} \},Q,\mu_h)  \, \otimes \,  \widehat{\bm{\mathcal{S}}}_3(\{\underline{n} \},\mu_h)  \big\rangle = \int_0^1 \!du \int_0^1 \!dv \, \big\langle \bm{\mathcal{H}}_3^{(1)}(u,v,Q,\mu_h) \widehat{\bm{\mathcal{S}}}_3(u,v,\mu_h) \big\rangle\,,
\end{align}
where we have defined $\widehat{\bm{\mathcal{S}}}_3(u,v,\mu_h) = \sum_{m=3}^\infty \bm{U}_{3m}(\{\underline{n}\},\mu_s,\mu_h) \,\hat{\otimes}\, \bm{1}$, which is the LL RG evolution or parton shower soft function. To implement this formula into a Monte Carlo framework, we will randomly generate $u$ and $v$ and then run the shower  $\widehat{\bm{\mathcal{S}}}_3(u,v,\mu_h)$ for the given configuration. There is, however, one complication, namely that the hard function is a distribution and can therefore not be integrated point by point. One way to solve this problem is to evaluate $\widehat{\bm{\mathcal{S}}}_3(u,v,\mu_h)$ on a grid, interpolate and then perform the integrations over $u$ and $v$. This works well because $\widehat{\bm{\mathcal{S}}}_3(u,v,\mu_h)$ is a smooth function of the angles as can be seen from Figure~\ref{fig:S3fun}. Note in particular that the limit $v\to 0$, in which both angles go to zero and the two Wilson lines become collinear, is completely smooth. In this limit the quark and gluon Wilson lines combine and produce the same radiation as a single quark Wilson line, encoded in the function $\widehat{\bm{\mathcal{S}}}_2$. The relation
\begin{equation}\label{eq:S3toS2}
\widehat{\bm{\mathcal{S}}}_3(u,v=0,\mu_h)=\widehat{\bm{\mathcal{S}}}_2(\mu_h)
\end{equation}
will lead to important simplifications below. In the right plot, we show the evolution time dependence of the soft function $\widehat{\bm{\mathcal{S}}}_3$ for fixed angles. One observes that the function falls off much faster when the hard partons approach the jet cone. In this configuration, more soft radiation exits the cone, explaining this suppression.

Interpolating the soft function $\widehat{\bm{\mathcal{S}}}_3$ gives accurate results, but is not efficient since the function depends on the phase-space constraints and thus needs to be recomputed when one changes the cone angle. It is much more natural to compute the convolution \eqref{H3S3} directly in the Monte Carlo code. The simplest way to implement the plus distributions in the hard function into the Monte Carlo is to use a slicing method. To explain it in a simple setting, let us for the moment only consider the $v$ dependence and forget about the variable $u$. Then the convolution \eqref{H3S3} takes the form 
\begin{align}\label{slicing}
\mathcal{H}_3^{(1)}  \, \otimes \,  \widehat{\mathcal{S}}_3 = \int_0^1 dv \left[ A\,  \delta(v)   + B(v) + \sum_{i=0}^1 C_i(v) \left( \frac{\ln^i v}{v} \right)_+\right] \widehat{ \mathcal{S}}_3(v),
\end{align}
where $B(v)$ represents a regular function. Thanks to relation \eqref{eq:S3toS2} the
 $A$ term can be combined with the LL parton shower result involving $\widehat{\bm{\mathcal{S}}}_2$ and the contribution from $B(v)$ can be computed by randomly generating $v$-values and running the shower for each chosen configuration. The slicing method introduces a lower cutoff $v_0$ into the plus distribution integrals $C_i(v)$ to ensure that $v$ can not go to zero. With the cutoff in place, we can integrate the subtraction term, e.g.
\begin{align}\label{slicingPlus}
\int_0^1  \frac{d v}{v} \left[ \widehat{ \mathcal{S}}_3(v) - \widehat{ \mathcal{S}}_2  \right] = \int_{v_0}^1 \frac{d v}{v} \widehat{ \mathcal{S}}_3(v) + \ln v_0 \, \widehat{ \mathcal{S}}_2 + \mathcal{O} (v_0),
\end{align}  
where one can use the same Monte Carlo method as for the $B(v)$ terms to simulate the first term with the collinear cutoff $v_0$, and then adds back the second  term which is given by the LL parton shower result, multiplied by a logarithm of the cutoff parameter. The $v_0$ dependence will cancel out between the two terms up to power corrections. The power corrections in the artificial parameter $v_0$ can be neglected as long as one chooses it small enough. The slicing method involves large cancellations between the two terms on the right-hand side of \eqref{slicingPlus}, so for numerical stability reasons one should not choose $v_0$ too small. These two opposing requirements make slicing methods delicate, but we compared to the result using the interpolated soft function $\widehat{\bm{\mathcal{S}}}_3$ and found good consistence. The cutoff independence is demonstrated in  Figure~\ref{fig:cutoff_dep} in Appendix~\ref{IJhs}. 

Up to now we have disregarded the $u$ dependence, but the Monte Carlo implementation of the full equations~\eqref{H3I}, \eqref{H3II} and \eqref{H3III} involves nothing beyond  the above discussion, except that we have to consider both integrations. As \eqref{eq:S3toS2} shows, the soft function becomes trivial for $v\to 0$ and we can combine all $\delta(v)$ dependent terms with the parton shower for $\widehat{\bm{\mathcal{S}}}_2$. We thus only need to apply the slicing method to the $\delta(u)\left(\ln^i v/v\right)_+$ and $(1/u)_+(1/v)_+$ terms. The corresponding cutoff dependent compensation terms are collected in Appendix~\ref{IJhs}. 

The final ingredient we need to implement is the one-loop soft function, which is defined as a sum over all dipoles
\begin{align}\label{eq:SmNLOa}
   \frac{\alpha_s}{4\pi}&\bm{\mathcal{S}}^{(1)}_m(\{\underline{n}\},Q_0,\e) =  \nno \\
  & - g_s^2\,\tilde{\mu}^{2\e} \sum_{(ij)}\,\bm{T}_{i,L}\cdot\bm{T}_{j,R} 
  \int\! \frac{d^{d}k}{(2\pi)^{d-1} }\delta(k^2)\theta(k^0) \frac{n_i\cdot n_j}{n_i\cdot k\,n_j\cdot k} \Theta_{\rm out}(n_k) \theta(Q_0 - E_k) \, ,
\end{align}
where the sum runs over all unordered pairs $(ij)$. In the large-$N_c$ limit only neighbouring legs give a contribution
\begin{equation}
{\bm T}_{i,L}\cdot{\bm T}_{j,R} \to - \frac{N_c}{2}\,\delta_{i,j\pm 1} \,.
\end{equation}
We evaluate the one-loop soft function numerically within our Monte Carlo code. It is well suited for this task since it generates emissions between neighbouring dipoles in an efficient way, by randomly choosing the rapidity $\hat y$ and azimuthal angle $\hat\phi$ of the emission in the COM (center-of-mass) frame of the emitting dipole $(n_i, n_j)$. Here and in the following, we will use hats to indicate kinematic quantities in the COM frame. Our hard function shower keeps emitting additional hard partons until one of them enters the veto region at which point it terminates. In our implementation, we use this last parton in the veto region to obtain the NLO correction to the soft function. At NLO, the renormalized soft function can be expressed as
\begin{align}\label{eq:SmNLOb}
\mathcal{S}_m^{(1)}(\{\underline{n}\},Q_0,\mu) = \frac{N_c}{2} \sum_{i,j=1}^m \delta_{i,j\pm 1} \int d\hat y \, \int_0^{2\pi} \frac{d\hat\phi}{2\pi} \left[ -4 \ln\frac{\mu}{Q_0} + 4\ln\frac{2\, |\sin\hat\phi|}{f_{ij}(\hat\phi,\hat y)} \right] \Theta_{\rm out}^{\rm lab}(\hat y,\hat\phi)\, ,
\end{align}
with $\Theta_{\rm out}^{\rm lab}(\hat y,\hat \phi)$ constraining soft radiation to be outside of the jet cone in the lab frame. In the Monte Carlo implementation, the factor in square brackets is a weight factor for the corresponding emission.  The auxiliary function $f_{ij}(\hat\phi,\hat y)$ connects the transverse momentum $\hat{k}_T$ in the COM frame to the energy $Q_0$ in the lab frame, $\hat{k}_T  f_{ij}(\hat \phi,\hat y) \leq Q_0$, and is given by $f_{ij}(\hat\phi,\hat y) = \frac{2}{M} \left(-\beta\cos\hat\phi + \cosh \hat y\right)$, where $M^2 = 2\, n_i \cdot n_j$ is the invariant ``mass'' of the dipole pair, and $\beta=\sqrt{1-M^2/4}$. The logarithm of $|\sin\hat\phi|$ arises from expanding the azimuthal angular integration in $\epsilon$, which is related to the  space-time dimension through $d=4-2\e$. A detailed derivation of expression \eqref{eq:SmNLOb} can be found in Appendix~\ref{IJhs}. 

While our slicing implementation of the hard function is simple but specific to the dijet processes and certainly not optimal, the above procedure to obtain the NLO soft function is simple, efficient and general. Compared to the LL parton shower code, including the one-loop soft function correction \eqref{eq:SmNLOb} yields
\begin{align}\label{eq:sigmaLLpSoft}
 \sum_{m=2}^\infty \big\langle  \bm{\mathcal{H}}_m(t) \,\hat{\otimes}\, \bm{\mathcal{S}}_m^{(1)}\big\rangle = \big\langle \bm{\mathcal{H}}_2(t)\bm{\mathcal{S}}_2^{(1)} + \int \frac{d\Omega_1}{4\pi} \bm{\mathcal{H}}_{3}(t) \bm{\mathcal{S}}_3^{(1)} +\int \frac{d\Omega_1}{4\pi}\!\int \frac{d\Omega_2}{4\pi}  \bm{\mathcal{H}}_{4}(t)\bm{\mathcal{S}}_4^{(1)} + \dots \big\rangle \,,
\end{align}
where one evolves the hard function from hard scale to soft scale and multiplies it with the soft function $\bm{\mathcal{S}}_m^{(1)}$ of the corresponding multiplicity. When running our Monte Carlo code we fill three histograms, one for the LL shower, one for the logarithmic part of \eqref{eq:SmNLOb} and one for the non-logarithmic part. Further details of the Monte Carlo algorithm, including the implementation of the one-loop soft function are given in  Appendix~\ref{IJmc}. 

The computer time needed to run the shower including the one-loop corrections depends on the maximum evolution time needed in the computation. For the interjet energy flow, we run the shower until $t=0.08$, corresponding to $\mu_s \approx 1\,{\rm GeV}$. For a collinear cutoff at $\eta_{\rm cut}=4$ ($\eta_{\rm cut}=5$) in the parton shower we then end up with about $15$ ($30$) hard partons per event on average. To resolve the peak region of the jet mass, discussed in the next section, we have to run to extremely low scales $\mu_s = 0.275\,{\rm GeV}$, corresponding to $t=0.3$, near the Landau pole at $\Lambda = 0.230\,{\rm GeV}$. At this scale, hundreds of partons are generated in each event and we need a few days of computer time on a cluster to obtain our numerical results, which will be presented in Section~\ref{sec:numerics} below.

\section{NLL$\rm'$ resummation for jet mass\label{sec:jetmass}}

Our second task is to perform the resummation for the jet mass distribution at electron-positron colliders. In contrast to the interjet energy flow, this observable suffers from soft-collinear double logarithms. These then constitute the LL results, while the non-global structure only arises at NLL. The resummation of jet mass including the leading non-global logarithms has been discussed in \cite{Dasgupta:2001sh,Banfi:2010pa,Dasgupta:2012hg,Becher:2016omr}.  At NLL level, the non-global logarithms yield a simple overall factor which multiplies the cross section. Beyond NLL this simple factorization does not hold anymore, and one needs to include the corrections piece by piece.\footnote{The recent paper \cite{Cal:2019hjc} on the jet shape includes one-loop corrections only for the global part, which corresponds to $m=1$ in \eqref{eq:factlightmassLap}, and does therefore not  reach full ${\rm NLL}'$  accuracy. Including the non-global structure would result in a factorization formula similar to (4.18) in \cite{Balsiger:2018ezi}.} The basic structure of the corrections is of course the same as for the interjet energy flow, see \eqref{LLp} and Figure~\ref{fig:InterJ_LLp}, and we therefore mainly focus on the differences to this case. In addition to the double logarithms, the most important new element is that the factorization arises in Laplace space. We use the same notation as \cite{Becher:2016omr}, where we presented NLL resummation results. For $\rm NLL'$ accuracy we need to keep one-loop matching corrections in the factorization formula \eqref{eq:factlightmassLap} and the theorem then reads
\begin{align}\label{NLLpLjet}
\tilde\sigma(\tau,\mu_h)  = &\sum_{i=q,\bar{q}} \sum_{m=1}^\infty\tilde{j}_i(\tau Q,\mu_h) \, \big\langle \bm{\mathcal{H}}^i_1(\{\underline{n}\}, Q,\mu_h) \otimes \bm{U}_{1m}(\{\underline{n}\},\mu_s,\mu_h) \, \hat \otimes  \, \bm{\mathcal{\widetilde{S}}}_m(\{\underline{n}\},\tau,\mu_s) \big\rangle \nno \\
& + \sum_{i=q,\bar{q},g} \sum_{m=2}^\infty \tilde{j}_i(\tau Q,\mu_h) \, \big\langle \bm{\mathcal{H}}^i_2(\{\underline{n}\}, Q,\mu_h) \otimes \bm{U}_{2m}(\{\underline{n}\}, \mu_s,\mu_h) \, \hat \otimes \, \bm{1} \big\rangle\,. 
\end{align}
In the first line we must include one-loop corrections for the quark jet function $\tilde{j}_q$, the hard function $\bm{\mathcal{H}}_1$ and soft functions $\bm{\mathcal{\widetilde{S}}}_m$. We do not include the $\mathcal{O}(\alpha_s^2)$ cross terms so that the first line turns into a sum of terms with the individual corrections. The hard function $\bm{\mathcal{H}}_2^i$ in the second line includes two hard partons in the right jet. Since it involves a power of $\alpha_s$ due to the hard emission, the remaining ingredients are only needed at LO. The second line also includes a gluon-jet contribution, for the case where the $q\bar{q}$ pair is in the right hemisphere. The one-loop hard functions are the same as for interjet energy flow, up to the different phase-space constraints. They are given in Appendix~\ref{LjetHS}.

In Laplace space, RG-evolution is multiplicative and we can factor out and exponentiate the double logarithms. Removing the double logarithmic part is important since our shower evolution, which also takes place in Laplace space, is purely soft. The subtraction of collinear contributions will also be needed for our numerical computation of the one-loop soft function. Using standard techniques introduced in \cite{Becher:2006nr}, we can perform the inversion to momentum space analytically at the end and write a momentum space result directly in terms of Laplace-space ingredients.

The anomalous dimension $\bm{\Gamma}^H$ in \eqref{eq:gammaOne} which drives the resummation of the logarithms in interjet energy flow \eqref{sigbarefinal} can be viewed in two ways: As the hard anomalous dimension, used to evolve the hard functions to the soft scale, or as the soft anomalous dimension  which evolves the soft functions to a higher scale. RG invariance of the cross section implies that the two evolutions must agree. The situation is more interesting for the light-jet mass \eqref{eq:factlightmassLap} which involves three ingredients. In this case RG invariance translates into the statement
\begin{equation}
\bm{\Gamma}^{H_i}_{lm}(\{\underline{n}\}, Q,\mu)= \bm{\Gamma}^{S_i}_{lm}(\{\underline{n}\}, \tau,\mu) + \Gamma^{J_i}(\tau Q,\mu) \delta_{lm}\,,
\end{equation}
where 
\begin{equation}
 \Gamma^{J_i}(\tau Q,\mu) = -2 C_i \gamma_{\rm cusp} \ln\!\left(\frac{\tau Q}{\mu^2}\right) + 2\gamma^{J_i}\,.
 \end{equation}
 The Casimir $C_i$ for the quark-jet channel is $C_q = C_F$, while the gluon configuration has $C_g = C_A$.
In our paper \cite{Becher:2016omr}, we have analyzed the one-loop soft anomalous dimension and found that it has the form
\begin{equation}
\bm{\Gamma}^{S_i}_{lm}(\{\underline{n}\}, \tau,\mu) =  2 C_i\, \gamma_{\rm cusp} \,\ln\! \left(\frac{\tau}{\mu} \right) \delta_{lm}  + \bm{\hat{\Gamma}}_{lm}(\{\underline{n}\})\,, \label{eq:sftGamma}
\end{equation}
where $ \bm{\hat{\Gamma}}_{lm}$ is a regular non-logarithmic anomalous dimension, which takes the same form as  \eqref{eq:gammaOne}, except for a subtraction to remove the collinear singularities, which give rise to the cusp piece in \eqref{eq:sftGamma}. The subtraction is achieved by  replacing the diagonal elements in  \eqref{eq:gammaOne} by  $\bm{V}_m \to \overline{\bm{V}}_m = \bm{V}_m - \bm{V}_0$, with
\begin{align}\label{subtraction}
\bm{V}_0 = V_0\, \bm{1} = -4\, C_i\,  \bm{1}  \int \frac { d \Omega \left( n _ { k } \right) } { 4 \pi }  \frac{\bar n \cdot n}{\bar n \cdot n_k \, n_k \cdot n} \Theta_{\rm L}(n_k),
\end{align}
where $\Theta_{\rm L}(n_k)$ ensures that the emission is in the left hemisphere with the light jet. The trivial color structure arises from color conservation
\begin{equation}
\sum_{i=1}^m \bm{T}_{0}\cdot  \bm{T}_{i} = - \bm{T}_{0}\cdot  \bm{T}_{0} = -C_i \,\bm{1}\,.
\end{equation}
Note that $V_0$ is equal to the one-loop result (real plus virtual) for the case where there is only one hard parton on the right, which then, by momentum conservation, flies along $n$. The subtraction therefore removes the ``global'' one-loop part of the soft anomalous dimension. After this, the Monte Carlo result no longer involves collinear singularities. As before we regularize the collinear singularities in the individual entries of $\bm{\hat{\Gamma}}$ using a cutoff. The parton shower algorithm of Dasgupta and Salam \cite{Dasgupta:2001sh} instead uses a veto algorithm to remove global logarithmic terms. Our  subtraction of the global piece has the advantage that our Monte Carlo weights are always positive. Let us also note that the role of the subtraction is to separate out the collinear singularities, so that the same subtraction can be used for any process with the same double logarithmic structure, i.e. also in cases with more complicated geometry, where we cannot analytically compute the one-loop function.

To make use of the separation of the anomalous dimension into two pieces, we now factor the soft function as
\begin{equation}
\widetilde{\bm{\mathcal{S}}}^i_m(\{\underline{n}\},\tau,\mu_s)  =\widetilde{S}_G^i(\tau,\mu_s) \bm{\hat{S}}_m^i(\{\underline{n}\},\tau,\mu_s) \,,
\end{equation}
with
\begin{align}\label{sGq}
\widetilde{S}_G^q \left( L_s = \ln\frac{\tau}{\mu_s},\mu_s \right) = 1 + \frac{\alpha_s}{4\pi} C_F\left( - 4 L_s^2 - \frac{\pi^2}{2} \right) + \mathcal{O}(\alpha_s^2). 
\end{align}
The splitting of the soft function into single and double logarithmic pieces is of course not unique. We have chosen the double-logarithmic ``global'' part $\widetilde{S}_G^i$ such that it includes the full one-loop result, so that the ``non-global'' remainder function $\bm{\hat{S}}_m^i$ starts at two loops for $m=1$ partons in the right hemisphere. For the gluon case, we only need the tree-level result $\widetilde{S}_G^g=1$ since the hard function for this channel is suppressed by $\alpha_s$.

The global piece fulfills a standard RG-evolution equation driven by the cusp piece of \eqref{eq:sftGamma} which can be immediately solved in Laplace space. Using the technique introduced in  \cite{Becher:2006nr}, the associated momentum-space solution takes the form
\begin{align}\label{SGrun}
S_G^i(\omega,\mu) = \exp\left[ 2 C_i S(\mu_s,\mu) \right] \widetilde{S}_G^i(\partial_{\eta_S},\mu_s) \frac{e^{-\gamma_E \eta_S}}{\Gamma(\eta_S)} \frac{1}{\omega} \left( \frac{\omega}{\mu_S} \right)^{\eta_S} ,
\end{align}
with  $\eta_S = 2 C_i A_{\gamma_{\rm cusp}}(\mu_s,\mu)$, where the logarithm $L_s$ has been replaced by a derivative operator with respect to $\eta_S$.

With the global function at hand, the  Monte Carlo simulation only needs to provide the  remainder $\bm{\hat{S}}_m^i$. Its single logarithmic RG-evolution is obtained by the subtracted parton shower described above and the one-loop correction for an $m$-parton configuration is given by
\begin{equation}\label{eq:oneSoftHemi}
 \bm{\hat{S}}_m^{i\,( 1)}(\{\underline{n}\},\tau,\mu_s) = 
 \bm{\widetilde S}_{ m }^{i\, (1)} \left( \{ \underline{n} \} , \tau , \mu _ { s } \right)  - \widetilde S_ { G }^{i \,(1)}( \tau,\mu_s)\, ,
\end{equation}
which, by construction, is free from collinear logarithms. We compute this difference in the large-$N_c$ limit by running the shower until it produces a parton in the left hemisphere, which is the veto region for the present case. The outside parton is the soft emission and we then compute the relevant one-loop weight factor precisely as in \eqref{eq:SmNLOb}. The form of the Laplace space soft function can be found in the appendix in \eqref{eq:SmJet}. When the emission arises from the first dipole, which involves the left parton along $n_0 = \bar{n}$, we subtract the global part. For the quark-jet channel the subtraction is given by
\begin{align}\label{eq:globalSubtr}
\widetilde{S}_G^{q\, (1)}(\tau,\mu) = \frac{N_c}{2} \int d\hat y \, \frac{d\hat\phi}{2\pi} \left[ -4 \ln\frac{\mu}{\tau} + 4\ln\frac{2\, |\sin\hat\phi|}{g_{0j}(\hat\phi,\hat y)} \right] \Theta_{\rm L}(\hat y,\hat\phi) X(\hat y,\hat\phi) \, ,
\end{align}
with a re-weighting factor
\begin{equation}
X (\hat{y},\hat{\phi}) = e^{2\hat{y}}/(e^{2\hat{y}} + \beta^2 - 2 e^{\hat{y}} \beta \cos\hat{\phi})\,.
\end{equation}
 The factor $X$ is simply the ratio of the radiator \eqref{eq:dipole} associated with the original $(\bar{n},n)$ dipole and the one of the dipole $(\bar{n},n_j)$ which emits the gluon and defines the frame in which $\hat{y}$ and $\hat{\phi}$ are generated. The subtraction removes the collinear divergence in the $(\bar{n},n_j)$ dipole and yields $\bm{\hat{S}}_m^q$. The function $g_{ij}$ in \eqref{eq:globalSubtr} relates the momentum component $\bar{n}\cdot k$ in the lab frame to the transverse momentum $\hat k_T$ in the COM frame of the dipole $(n_i,n_j)$, analogously to the function $f_{ij}$ in \eqref{eq:SmNLOb}. Its explicit form is given in the appendix in \eqref{eq:gfun}.

The final ingredients in \eqref{NLLpLjet} are the one-loop jet functions, which are well known. In Laplace space, the one-loop jet function is given by
\begin{align}
\tilde{j}_i\left(L_j=\ln\frac{Q\tau}{\mu^2},\mu\right) = 1 + \frac{\alpha_s}{4\pi} \left( C_i \gamma^{\rm cusp}_0 \frac{L_j^2}{2} + \gamma^{J_i}_0 L_j + c_1^{J_i} \right), 
\end{align}
which translates to the momentum-space result  \cite{Becher:2006nr}
\begin{align}
J_i(p^2,\mu) = \exp\left[ -4 C_i S(\mu_j,\mu) + 2 A_{\gamma^{J_i}}(\mu_j,\mu) \right] \tilde{j}_i(\partial_{\eta_J},\mu_j) \frac{e^{-\gamma_E \eta_J}}{\Gamma(\eta_J)} \frac{1}{p^2} \left( \frac{p^2}{\mu_j^2} \right)^{\eta_J}, 
\end{align}
with $\eta_J = 2 C_i A_{\gamma_{\rm cusp}}(\mu_j,\mu)$. The relevant expressions for the ingredients are listed in Appendix \ref{adapp}. Combining the global soft function with the jet function, we obtain
\begin{align}\label{eq:resummedRes}
\Sigma_i(\rho) = & \, Q^2 \int_0^\rho d\rho' \int_0^{Q \rho'} d\omega J_i(Q^2\rho' - Q\omega,\mu_h) S_G^i(\omega,\mu_h) \nno \\
= & \exp \left[ 2 C_i S \left( \mu _ { s } , \mu _ { h } \right) - 4 C_i S \left( \mu _ { j } , \mu _ { h } \right) + 2 A _ { \gamma ^ { J_i } } \left( \mu _ { j } , \mu _ { h } \right) \right] \tilde{j}_i(\partial_\eta,\mu_j) \widetilde{S}^i_G(\partial_{\eta_S},\mu_s ) \nno \\
&\times \frac { e ^ { - \gamma _ { E } \eta } } { \Gamma ( \eta + 1 ) } \left( \frac { Q ^ { 2 } \rho  } { \mu _ { j } ^ { 2 } } \right) ^ { \eta } \left( \frac { Q \mu _ { s } } { \mu _ { j } ^ { 2 } } \right) ^ { - \eta _ { S } },
\end{align}
where we define $\eta = \eta_J + \eta_S$. The full result is obtained after combining this with the subtracted shower evolution, the hard functions and the one-loop soft  correction \eqref{eq:oneSoftHemi}. To implement this expression in practice, we run the shower, tabulate the results for the individual contributions to \eqref{NLLpLjet} and then replace the global function $\widetilde{S}_G^i(\omega,\mu_h)$ in \eqref{eq:resummedRes} by the full result which includes the hard functions,  evolution and one-loop corrections.

Up to NNLL, the integrated heavy-jet mass distribution is obtained as
\begin{equation}
\frac{1}{\sigma_0}  \int_0^{\rho_h} d \bar \rho_h \frac{d\sigma}{d \bar \rho_h} = H_2(Q^2,\mu_h)\left[ \Sigma_q(\rho_h) \right]^2\,.
\end{equation}
Using this result and relation \eqref{eq:jetmassrel} one obtains the light-jet mass.

\section{Numerical results\label{sec:numerics}}

In this section we will present numerical results, first for the interjet energy flow, then for the jet mass. For our plots, we work with $Q=M_Z$ and $\alpha_s(M_Z)=0.1181$, and use two-loop $\alpha_s(\mu)$ running with $n_f=5$ quark flavors. To our knowledge, no measurements are available for the interjet energy flow, but we will compare our results for the jet mass to {\sc LEP} measurements by {\sc ALEPH} \cite{Buskulic:1992hq}. 

\subsection{Interjet energy flow}

\begin{figure}[t]
\centering
 \includegraphics[width=0.45\textwidth]{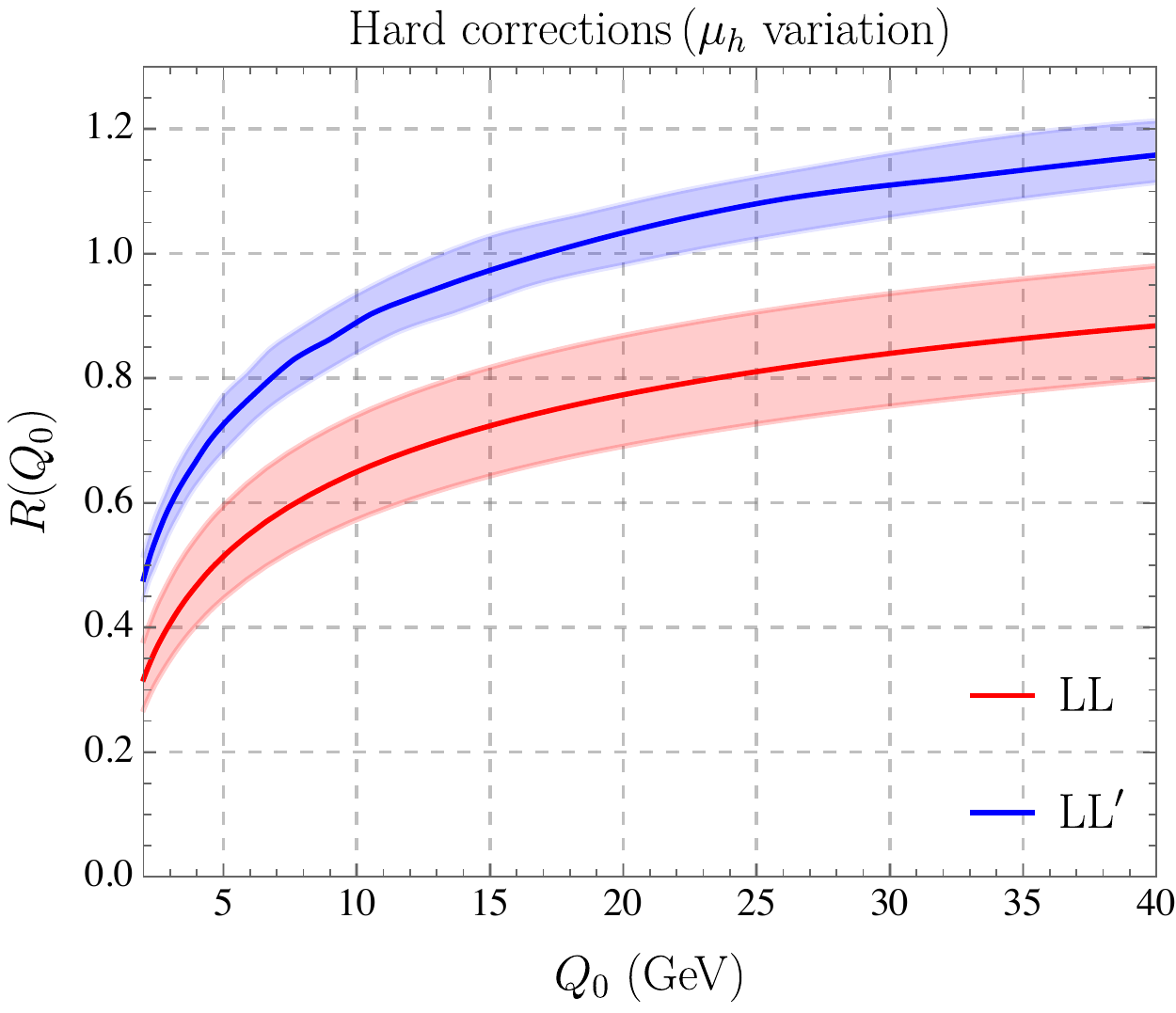}~~~~~
 \includegraphics[width=0.45\textwidth]{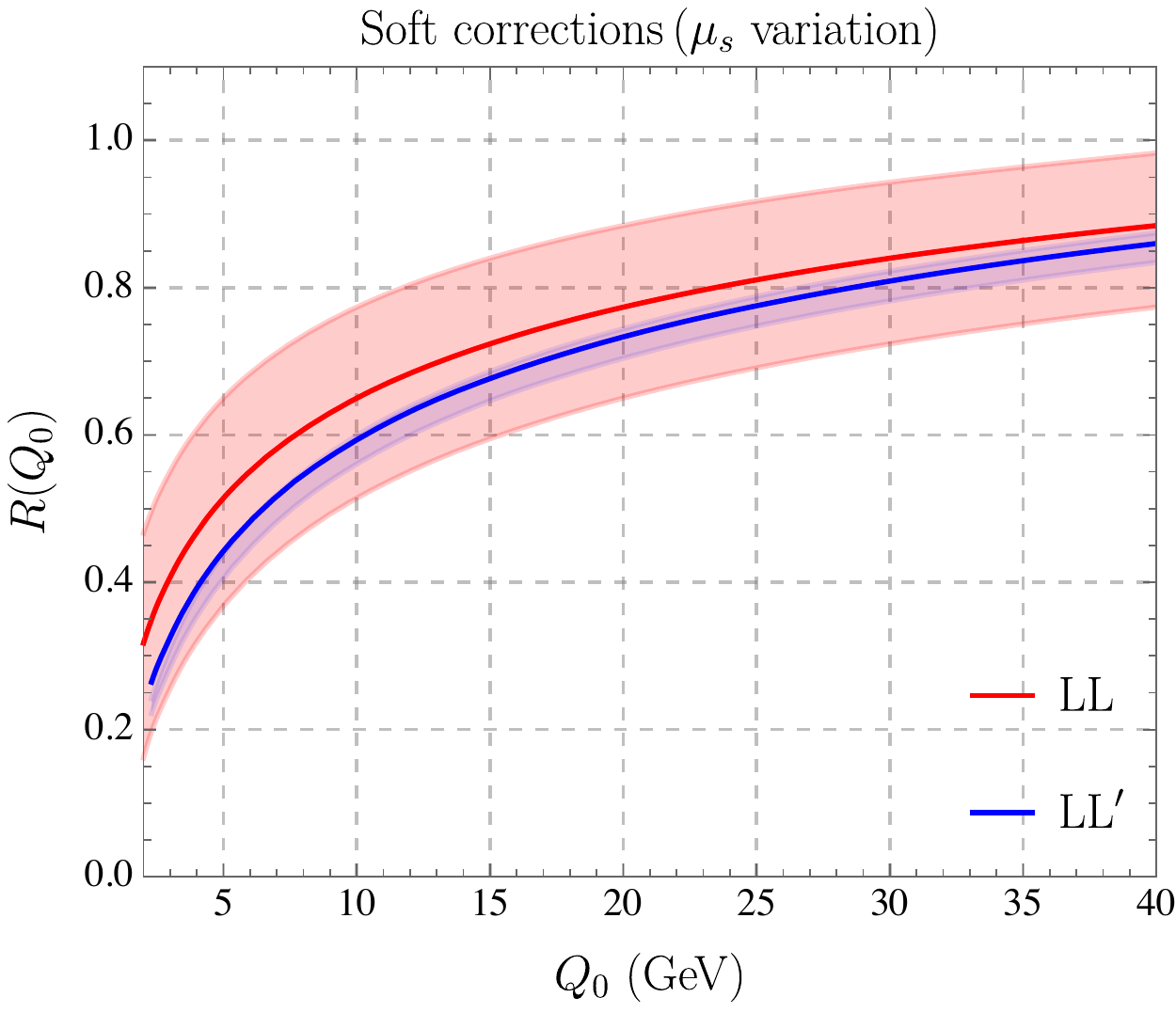}
 \caption{ Left panel: Hard function corrections, with bands arising from hard scale variation. Right panel: Soft function corrections, with bands from soft scale variation. \label{fig:InterJscale}}
 \end{figure}

For our numerical discussion we choose jet cone size parameter as $\alpha= \pi/3$. This is equivalent to $\delta = \tan \frac{\alpha}{2} = 1/\sqrt{3}$, or rapidity gap size $\Delta y = - \ln \delta^2 \approx 1.1$. We want to avoid small cone angles, or equivalently large rapidity gaps, in order not to have to deal with large collinear logarithms. In our plots we show the gap fraction
\begin{equation}
R(Q_0) = \frac{1}{\sigma_{\rm tot}} \sigma(Q, Q_0) \equiv  \int_0^{Q_0} dE_s\, \frac{1}{\sigma_{\rm tot}} \frac{d\sigma}{dE_s}\,,
\end{equation}
which is the fraction of events in which the soft radiation outside the jets has an energy $E_s$ below the cutoff $Q_0$. By definition, the amount of energy in the gap must be below $Q/2$, otherwise the thrust axis, which defines our jet axis, would flip. The fixed order result is therefore $R(Q_0=Q/2)=1$ at any order in perturbation theory. The $\mathcal{O}(\alpha_s^0)$ result with just two back-to-back partons is of course $R(Q_0)=1$, a nontrivial $Q_0$ dependence only arises at $\mathcal{O}(\alpha_s)$ when the third parton is inside the gap. We will refer to the $\mathcal{O}(\alpha_s)$ result as LO.

As a first step, let us check the size of the individual corrections and investigate whether the scale dependence is reduced after including them.  In Figure~\ref{fig:InterJscale} we show the hard and soft corrections separately and then plot the scale bands from varying the associated scales by a factor two around their default values $\mu_h = Q$ and $\mu_s=Q_0$. Compared to the LL scale bands shown in red, the scale dependence is reduced in both cases after including the corrections. We observe that the hard corrections are quite significant and positive, while the soft corrections are moderate and negative. The hard corrections have two sources, virtual corrections to $\bm{\mathcal{H}}_2$ and real emission contributions encoded in $\bm{\mathcal{H}}_3$. The first of these is just a constant factor multiplying the LL result, while the second one comes together with the higher soft function $\bm{\mathcal{S}}_3$. Both corrections are positive. At high values of $Q_0$ the three parton contribution from $\bm{\mathcal{H}}_3$ is about twice as large as the one from the one-loop correction to $\bm{\mathcal{H}}_2$ and it becomes more dominant at smaller values.

It is clear that the large hard function corrections at $Q_0 \lesssim Q/2$ must be compensated by terms which are power suppressed in $Q_0/Q$ and are not captured by the resummation based on the factorization formula \eqref{sigbarefinal}, which arises in the limit $Q_0 \to 0$. One can obtain these power suppressed terms by matching to the fixed-order result. More precisely, one adds to the resummed result the fixed-order prediction minus its expansion around $Q_0$. The subtraction removes the terms which are already included in the resummation. These power suppressed matching terms can be obtained as
\begin{equation}
\Delta R(Q_0) = \int_{0}^{Q_0} dE_s\, \frac{1}{\sigma_{\rm tot}} \left( \frac{d\sigma}{dE_s} - \left. \frac{d\sigma}{dE_s}\right|_{E_s \to 0} \right )\,.
\end{equation}
To evaluate this integral, one computes the cross section to find a parton inside the gap and subtracts from it its soft limit. The subtraction eliminates the virtual contributions and leads to a finite integral, which one can evaluate numerically. However, even after the matching to the fixed order result, the resummed result does not yet tend to $R(Q_0)=1$ for $Q_0 \to Q/2$ because we resum logarithms of $\mu_s /\mu_h \to 1/2$  for $\mu_s \approx Q_0$ and $\mu_h = Q$. To switch off the resummation, one can choose the soft scale in such a way that it approaches the hard scale $\mu_h$ as $Q_0 \to Q_{\rm max} = Q/2$. 
\begin{figure}[t]
 \centering
 \includegraphics[width=0.45\textwidth]{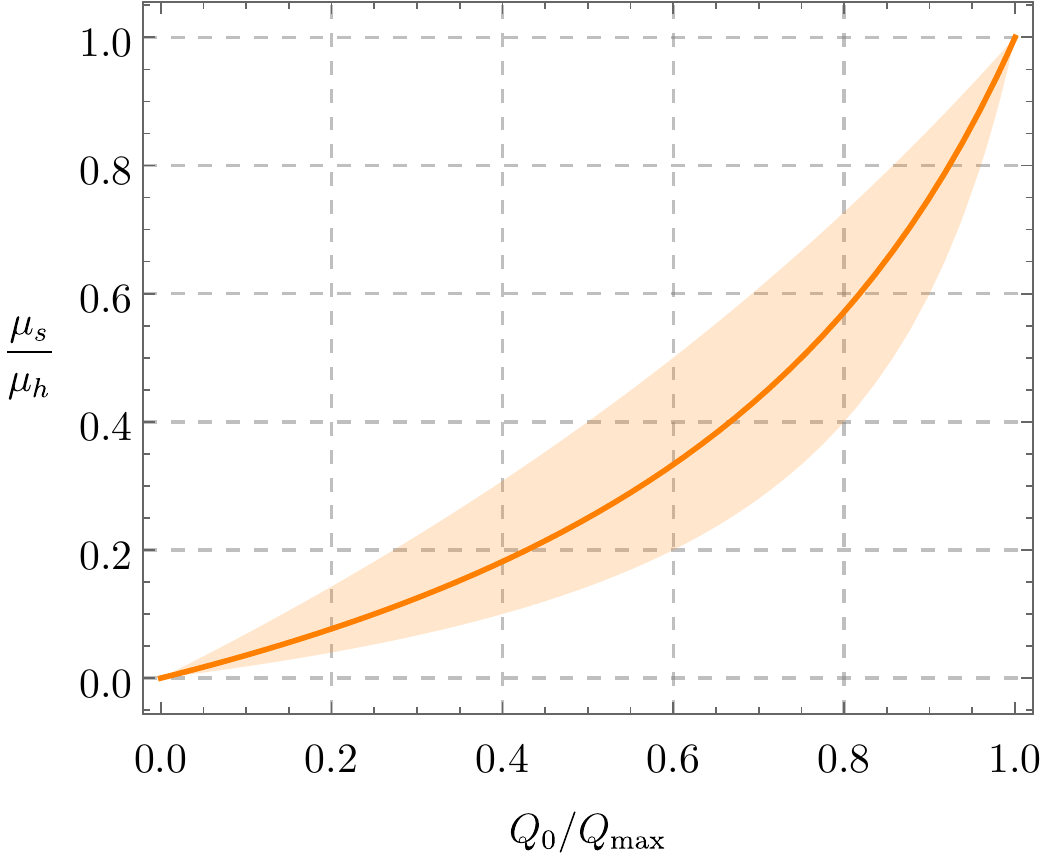}
 ~~~~~ \includegraphics[width=0.45\textwidth]{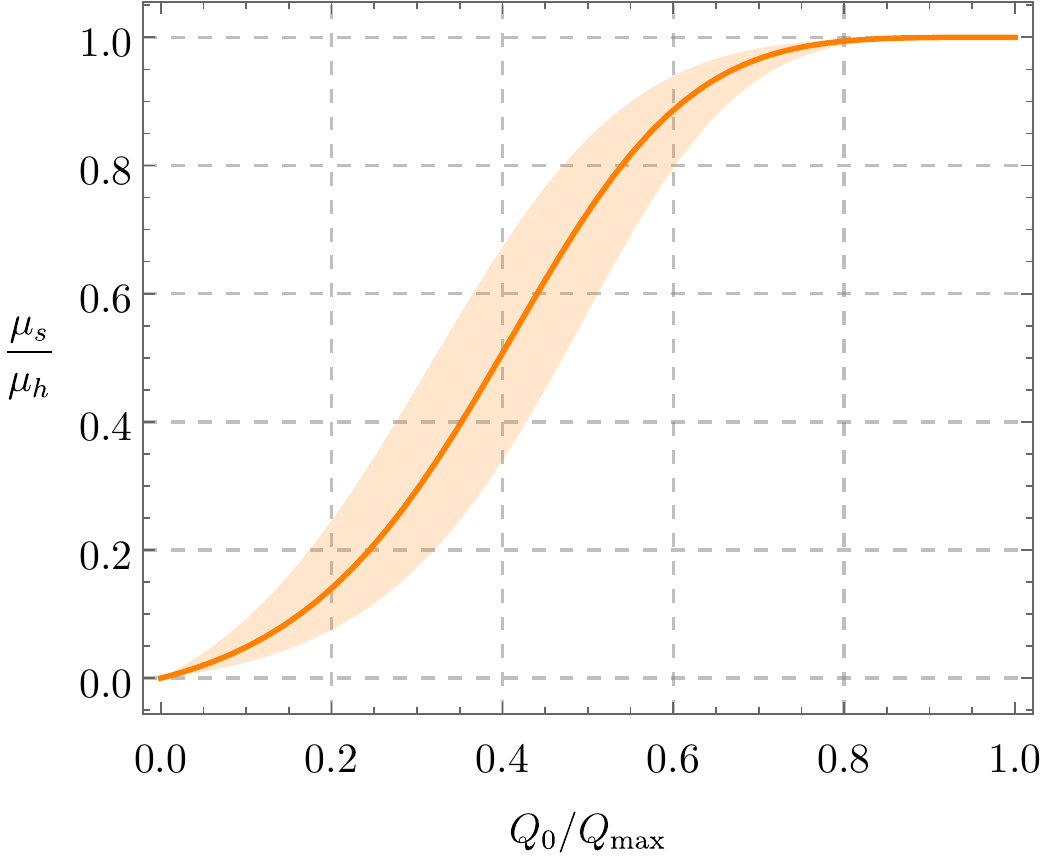}
 \caption{The profile function \eqref{profile} for $n=1$ (left plot) and $n=4$ (right plot). We will use the $n=4$ function as our default choice. \label{fig:profiles}}
 \end{figure}  
This can be achieved, for example with a {\em profile function} \cite{Abbate:2010xh} of the form
\begin{equation}\label{profile}
\mu_s(Q_0) = \frac{x_s \, Q_0}{1 + \frac{x_s Q_0}{\mu_h} + \sum_{i=1}^n c_i\, \Big(\frac{Q_0}{Q_{\rm max}}\Big)^i}\,,
\end{equation}
where $x_s=1$ corresponds to the default choice and the scale bands can be obtained by varying the parameter $x_s$ by a factor two. For low values 
of $Q_0$, this reduces to the standard choice $\mu_s(Q_0) =Q_0 x_s$. The power suppressed term in the denominator are chosen to switch off the resummation at the endpoint $Q_0 = Q_{\rm max}$, similarly to what is usually achieved through a modification of the logarithms in traditional resummation. The simplest choice for \eqref{profile} is $n=1$ and $c_1 = -1$, but we observe that the approach to fixed order is relatively slow. To make it faster, we choose $n=4$ and impose that the first three derivatives at the end-point vanish, explicitly $c_1 = -4$, $c_2 = 6$, $c_3 =  -4$, $c_4 = 1$.  We plot the two different profile functions in Figure~\ref{fig:profiles} and will use $n=4$ as the default in our numerical implementation. The choice of the profile function affects the resummation of power-suppressed contributions. If the shape is important, one should of course compute, or even resum, the power corrections to resolve the difference. The first step would be to include the matching up to NNLO, which would in principle be possible since the fixed-order results are available \cite{GehrmannDeRidder:2007hr,Weinzierl:2009ms,DelDuca:2016csb}. In practice it would require some effort since we would need to compute the fixed-order expansion of our results (including the shower).

\begin{figure}[t]
 \center
 \begin{tabular}{cc}
 \includegraphics[width=.45\textwidth]{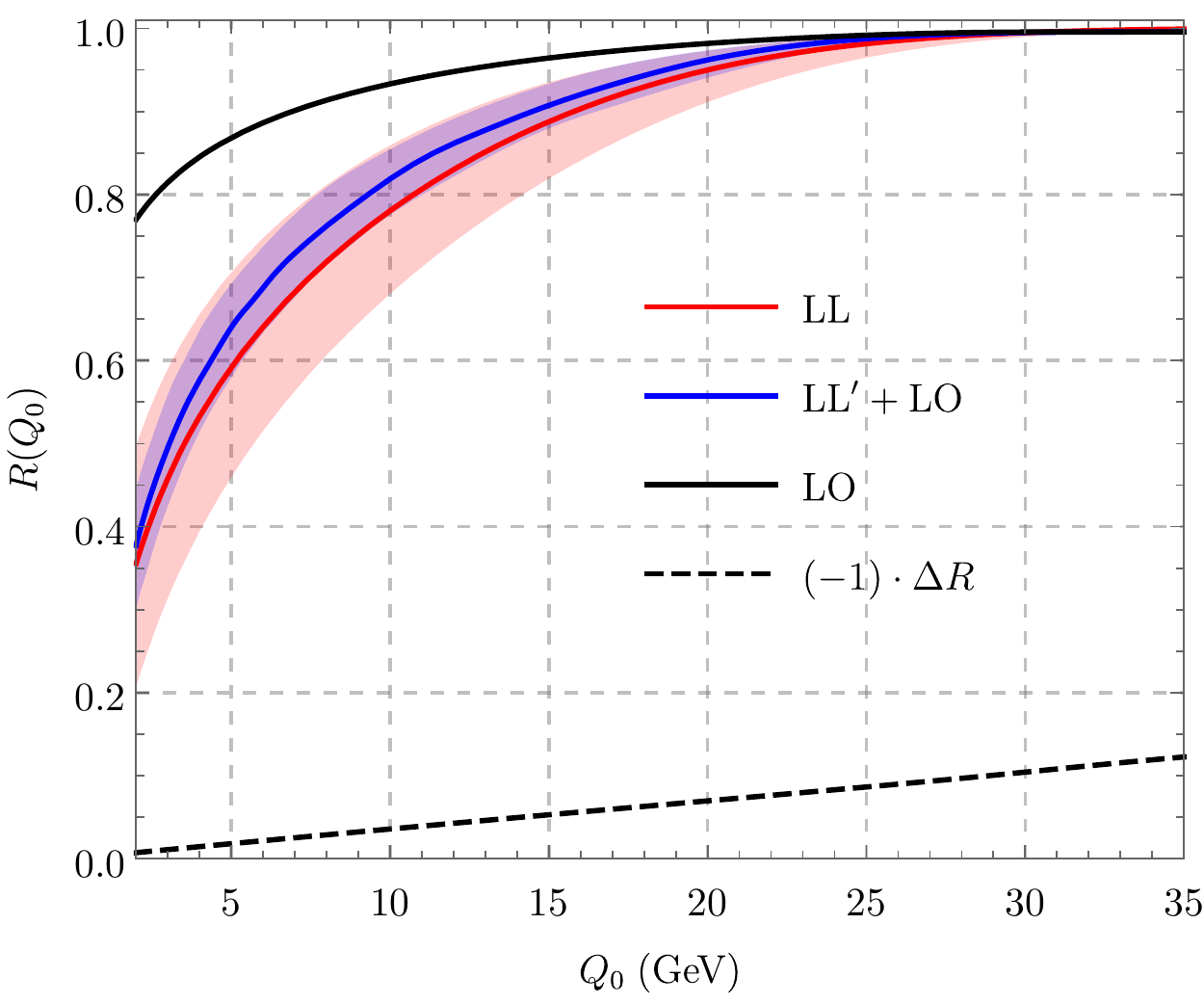} & \includegraphics[width=.45\textwidth]{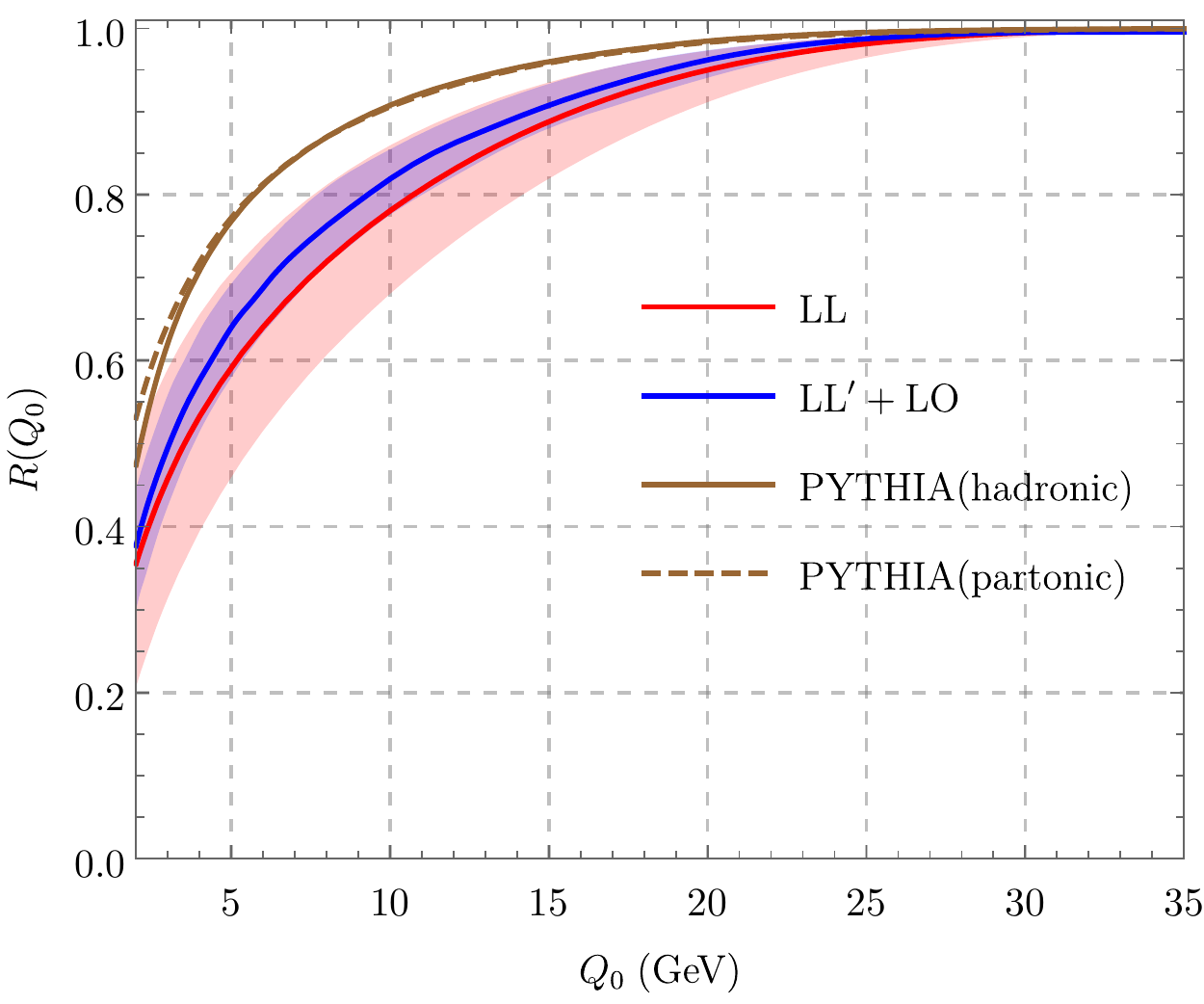}
 \end{tabular}
 \caption{Comparison of our results  for the interjet energy flow to fixed order (left plot) and to {\sc PYTHIA} (right plot).\label{fig:InterJFig}}
 \end{figure}

In Figure~\ref{fig:InterJFig}, we show an improved numerical result which includes the matching correction $\Delta R(Q_0)$, shown as a black dotted line, and uses the scale choice \eqref{profile} to switch off the resummation at the end-point. The matching correction is negative and compensates the large hard corrections near the end-point. The ${\rm LL}'$ corrections lead to a larger gap fraction $R(Q_0)$. As mentioned earlier, there is unfortunately no experimental data to which we can compare our results, but we compare to {\sc PYTHIA}  \cite{Sjostrand:2007gs}. While the two results are similar at very low $Q_0$, {\sc PYTHIA} is higher at intermediate values. We remind the reader, that the intermediate values heavily depend on the profile function used to switch off the resummation. 

%\begin{figure}[t]
% \center
% \includegraphics[width=.65\textwidth]{InterJFig.pdf}
% \caption{LL$'$+LO results for the inter-jet energy flow (blue curve). The scale uncertainty band is obtained from scale variations as explained in the text. The red band is the LL results, and black dashed-line is LO matching correction. \label{fig:InterJFig}}
% \end{figure}

\subsection{Jet mass}

\begin{figure}[t]
 \center
\hspace{-1cm} \includegraphics[width=1\textwidth]{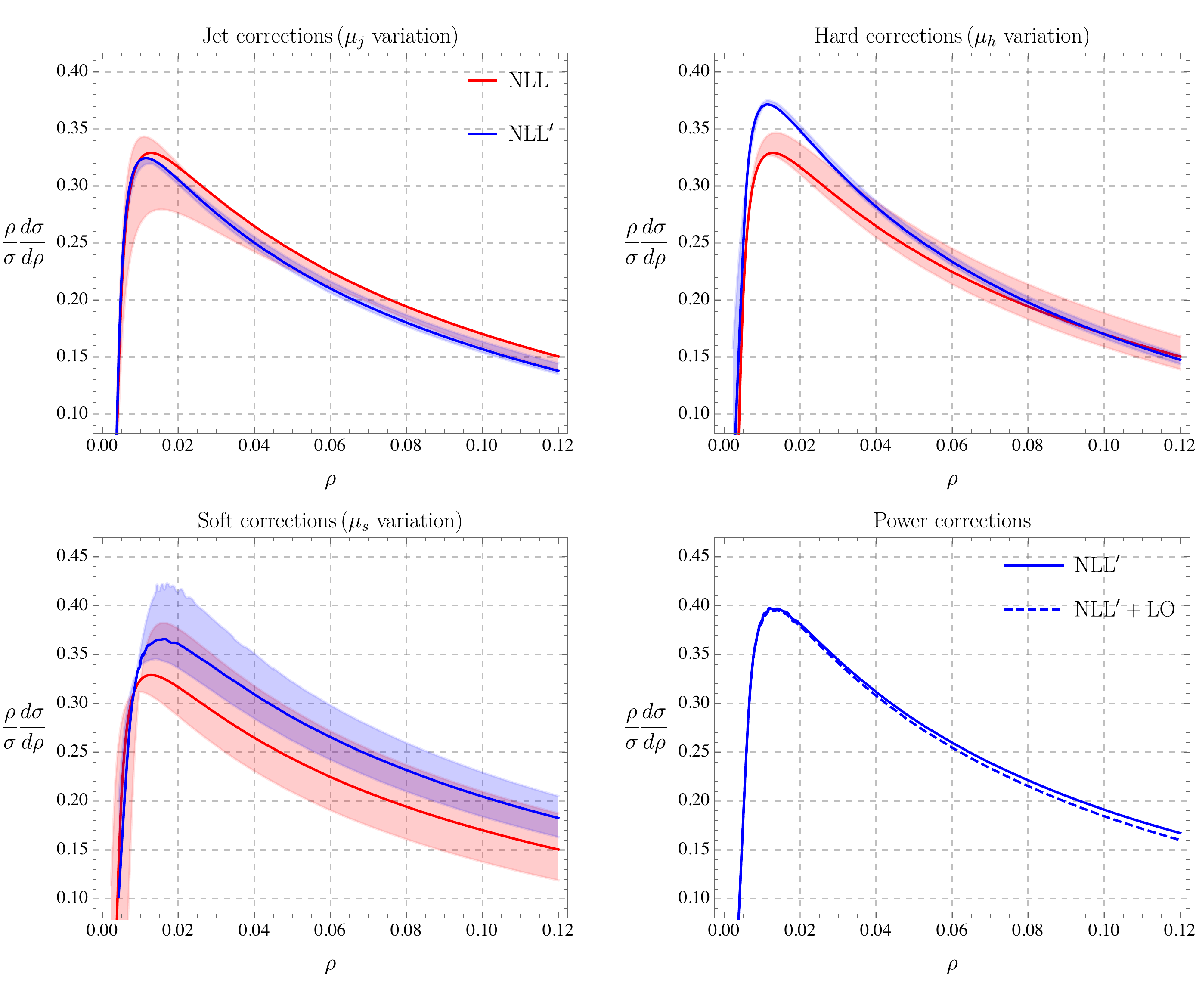}
 \caption{$\rm{NLL'}$ corrections from the jet, hard and soft functions and their scale uncertainties.  Each band comes from varying the scale associated with the correction by a factor of two around the default value. In the last plot we show LO power corrections from the fixed-order computation. We have multiplied the distributions by $\rho$ in order to make the results at larger $\rho$ visible.
 \label{fig:LeJetmuFig}}
\end{figure}

Let us now turn to the jet mass $\rho$. For interjet energy flow, we considered the integrated cross section, i.e. all events with energy in the gap below the veto, while we will look at the differential spectrum in the present case, since this is what was measured by the LEP experiments. We will however compute the spectrum by taking the derivative of the integrated cross section, which has the advantage that the spectrum is correctly normalized if the resummed prediction for the integrated cross section matches the fixed-order result at large $\rho$.

As a first step, we again separately plot the different ingredients and their scale dependence in Figure~\ref{fig:LeJetmuFig}. In the first three plots we compare NLL to $\rm{NLL'}$ with corrections from the jet, hard and soft functions. The red bands are the NLL result with scale variation, where we vary either the jet, hard or soft scale by a factor of two around the default values $\mu_h\sim Q$, $\mu_j \sim \sqrt{\rho}\,Q$ and $\mu_s \sim \rho\, Q$. The blue curves show contributions at $\rm NLL'$ accuracy from one of the three ingredients with its associated scale variation. Obviously, the scale dependence is strongly reduced from NLL to $\rm NLL'$ for jet and hard corrections. The soft scale dependence, on the other hand, is only modestly reduced after including one-loop soft function corrections.  The  scale bands mostly overlap with each other, which indicates that perturbative convergence is reasonably good in all the three cases. 

In the last plot of Figure~\ref{fig:LeJetmuFig} we show the effect of adding the $\mathcal{O}(\alpha_s)$ power corrections to the $\rm NLL'$ results. The LO power corrections for the heavy-jet mass  are known analytically and given in Appendix~\ref{adapp}. They are the same as for thrust, because the three-parton results for jet mass and thrust agree. Since the light-jet mass vanishes at $\mathcal{O}(\alpha_s)$, we can immediately also obtain the LO power corrections for the jet mass distribution. From the plot, we observe that the difference between $\rm NLL'$ and $\rm NLL' + LO$ is very small, and that the contributions from power corrections will reduce the resummed result in the large jet mass region. In order to reproduce the full fixed order result, we use $C_F = 4/3$ instead of the strict large-$N_c$ value $C_F = 3/2$ for the hard, jet and soft one-loop corrections in the resummed results. We also use the exact color factors in the evolution factors of the global part \eqref{eq:resummedRes}.

The end-point of the jet mass distribution is at $\rho_{\rm max} = 1/3$ at  $\mathcal{O}(\alpha_s)$, corresponding to a symmetrical configuration of the three partons. We will work with the same profile function \eqref{profile} to switch off the higher-order terms at the end point. To adapt it to the present case, we set $Q_0 = \rho\, Q$ and $Q_{\rm max} = Q/3$. For simplicity, we will adopt the canonical value $\mu_j =\sqrt{\mu_s \, \mu_h}$ in the following and only indirectly vary the jet scale through the variations of $\mu_s$ and $\mu_h$, which we vary independently by a factor of two around their default values.

\begin{figure}[t]
 \centering
 \includegraphics[width=0.45\textwidth]{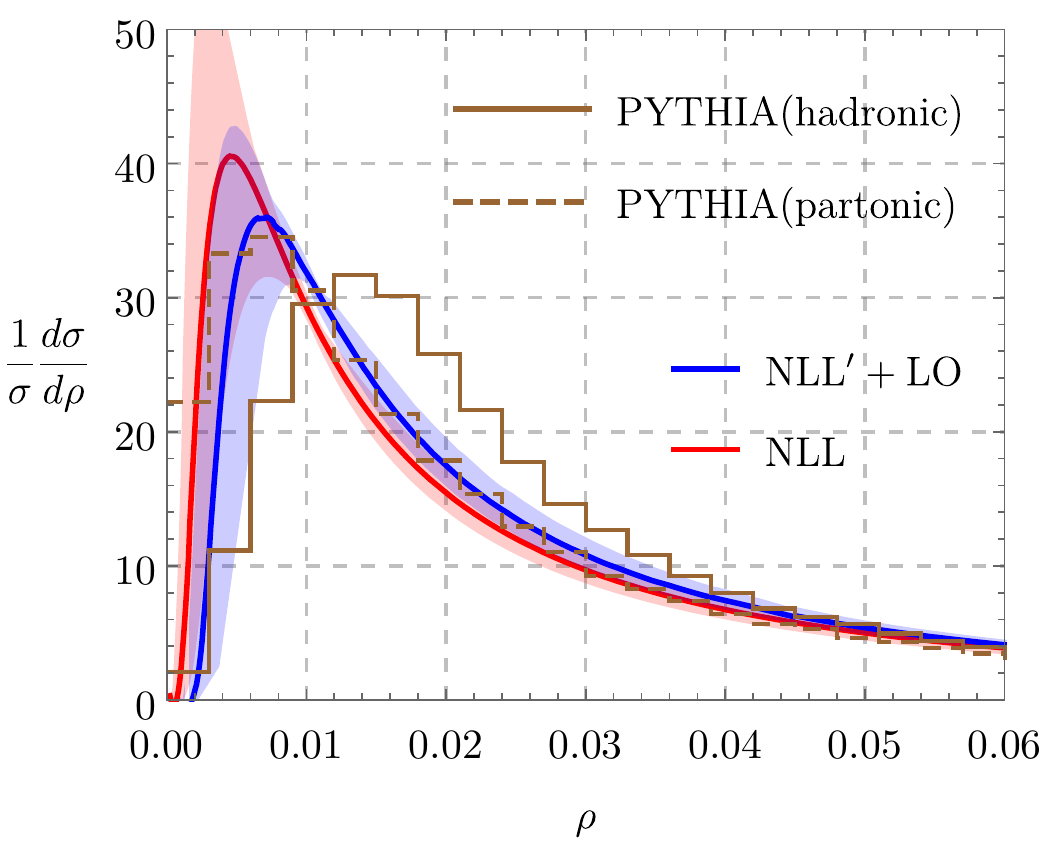} ~~~~~ \includegraphics[width=0.45\textwidth]{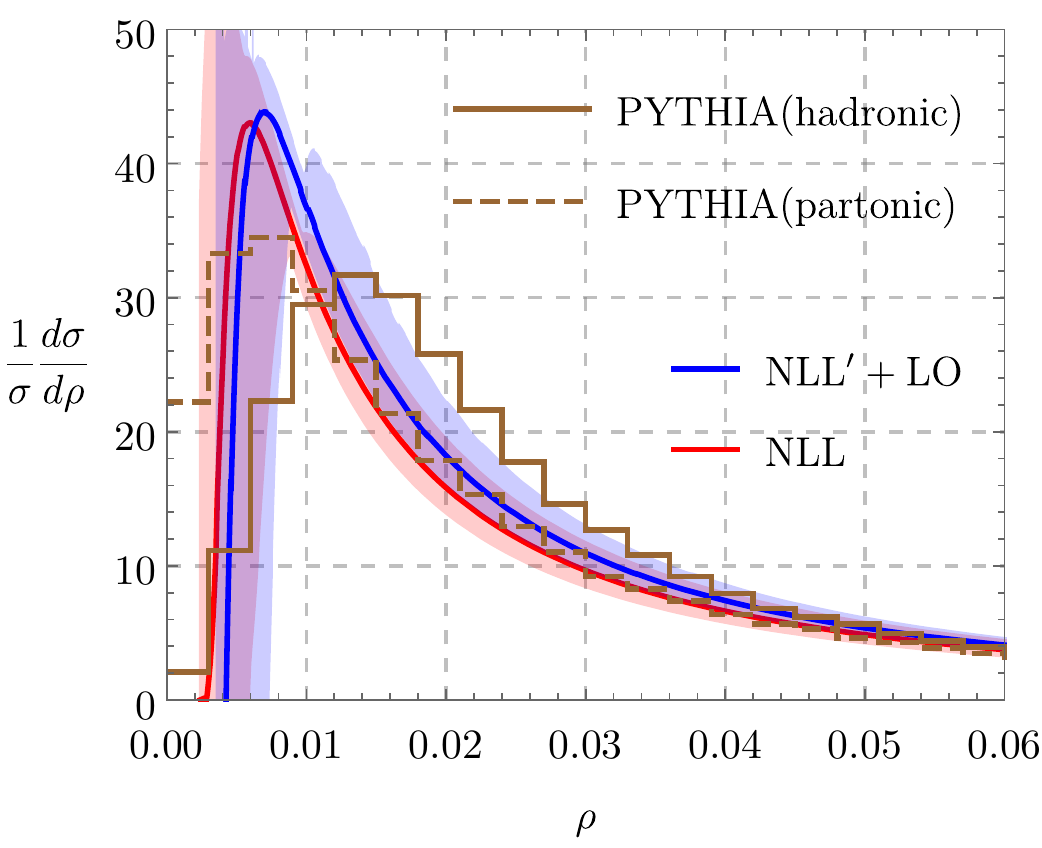}
 \caption{Jet-mass distribution compared to {\sc PYTHIA} results. On the left side we plot our  default result, based on using the profile scale \eqref{profile} and exponentiating the matching corrections. On the right-hand side, we do not perform these modifications such that we get a negative cross section at low $\rho$ and hit the Landau pole at a nonzero $\rho$. \label{fig:Pythia}}
 \end{figure}

At very low values of $\rho$, the scale $\mu_s(Q_0)$ hits the Landau pole at $\Lambda=0.23\,{\rm GeV}$. Near the pole the soft corrections become large and negative, resulting in a negative cross section. To avoid this unphysical behaviour, we replace $\mu_s(Q_0)\to \mu_s(Q_0) +\Lambda$ so that the pole occurs at $\rho=0$. We also exponentiate the hard, jet and soft corrections to avoid the negative cross section. In the left plot of Figure~\ref{fig:Pythia} we show our result for the jet mass distribution after these modifications. In the right plot, we show the result with $\mu_s(Q_0) = \rho\, Q$ and without exponentiation. We observe that the soft scale dependence changes sign at a point to the right of the peak. In this region the soft scale dependence becomes very small. With the modifications in $\mu_s$, we end up with quite small scale bands to the right of the peak, which are likely not an accurate characterization of the true uncertainties. The ${\rm NLL}'$ peak in the right-hand plot is quite a bit higher because the cross section becomes negative below $\rho = 0.004$ and our distributions are by construction normalized. An important feature of our result is that peak occurs at a very low value $\rho \approx 0.006$, which corresponds to $\mu_s \approx 0.5\,{\rm GeV}$ so that the peak region is strongly affected by nonperturbative effects. In Figure~\ref{fig:Pythia} we also show the {\sc PYTHIA} \cite{Sjostrand:2007gs} results, both on the parton level (dashed lines) and including hadronisation. The hadronisation effects shift the peak to the right by about $\Delta\rho \approx 0.006$, in accordance to what one expects from non-perturbative effects in the soft functions \cite{Korchemsky:1998ev,Korchemsky:1999kt}. The parton-level  {\sc PYTHIA} result is quite close to the ${\rm NLL}'$ result.

\begin{figure}[t]
 \centering
 \includegraphics[width=0.45\textwidth]{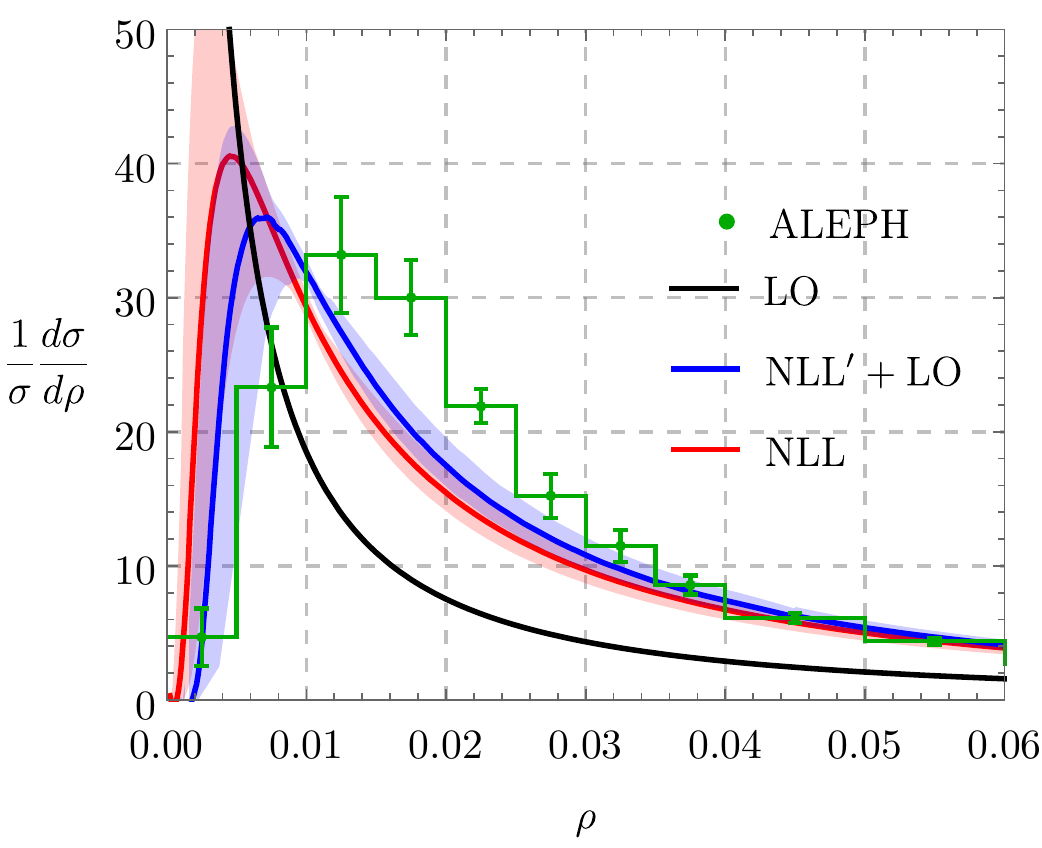} ~~~~~ \includegraphics[width=0.45\textwidth]{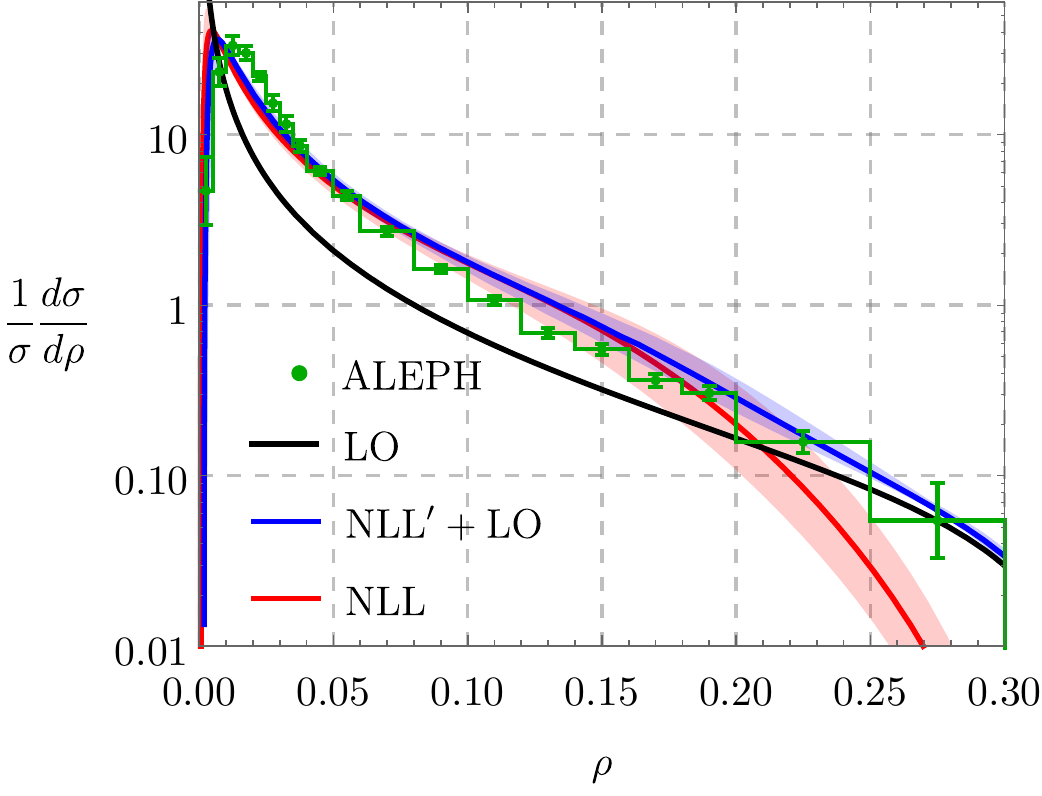}
 \caption{Jet-mass distribution and comparison to {\sc ALEPH} data \cite{Buskulic:1992hq} (green dots with error bars). The black curve represents the LO prediction for jet mass, where its analytical expression is given in \eqref{eq:LOjetmass}.  The red curve is the NLL resummation result and the band is from scale variation. The blue curve corresponds to $\rm NLL' + LO$ results, in which we switched off resummation effects at large $\rho$ using \eqref{profile}. \label{fig:LeftJet_NLLp}}
 \end{figure}

In Figure~\ref{fig:LeftJet_NLLp} we compare the $\rm NLL'+LO$ jet mass distribution with {\sc ALEPH} results \cite{Buskulic:1992hq}, obtained by combining their measurements for the light-jet and the heavy-jet mass using \eqref{eq:jetmassrel} and adding the uncertainties on the individual measurements in quadrature. One immediately sees that the experimental peak shifted to the right from non-perturbative effects and the shift is compatible with the {\sc PYTHIA} hadronization result. We also observe that the jet mass distribution falls off quite rapidly and to make the region of larger $\rho$ visible, we include also a logarithmic plot in Figure~\ref{fig:LeftJet_NLLp}. The plot also illustrates what motivated the profile function \eqref{profile} with $n=4$. The choices ensures that we start switching off the resummation fairly quickly about half-way to the endpoint and go over to the fixed-order result. The plots show that, compared the LO fixed-order result, resummation greatly improved the description of the experimental data. On the other hand there is  --- if at all --- only a relatively narrow region in $\rho$ in which both higher-order power corrections and non-perturbative corrections are small.

\begin{figure}[t]
\centering
\begin{tabular}{cc}
\includegraphics[width=0.43\textwidth]{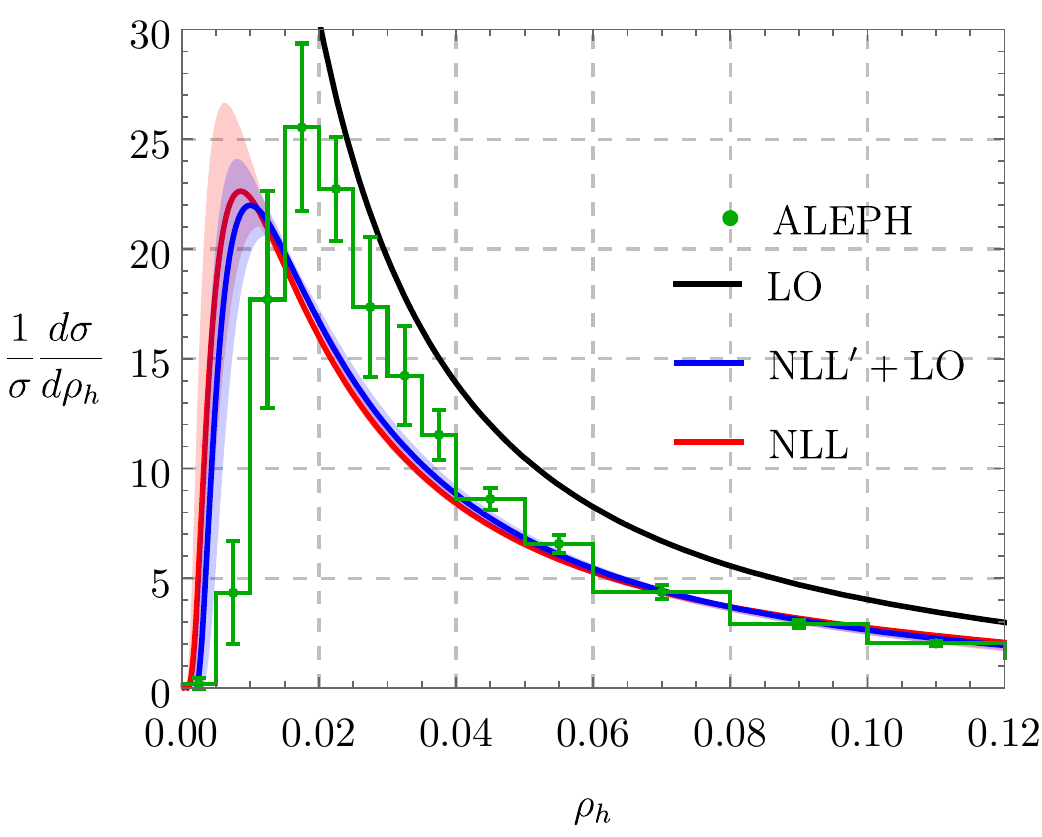} & \includegraphics[width=0.45\textwidth]{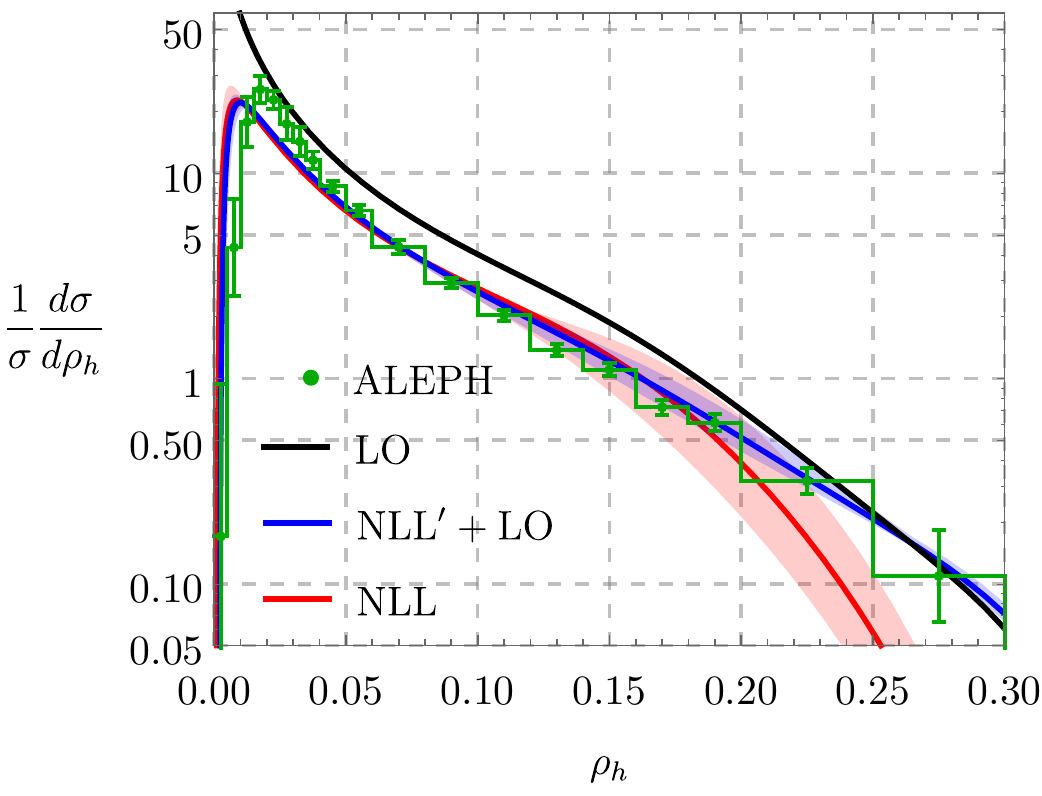} \\
\includegraphics[width=0.43\textwidth]{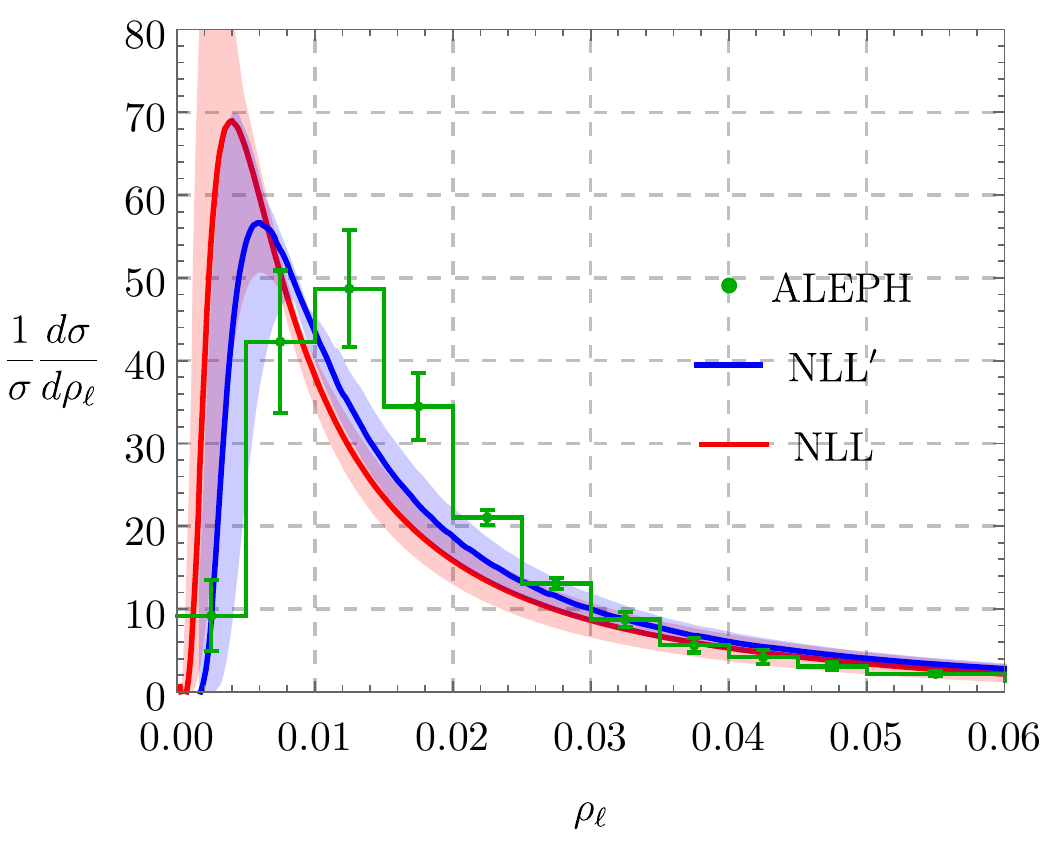} & \includegraphics[width=0.45\textwidth]{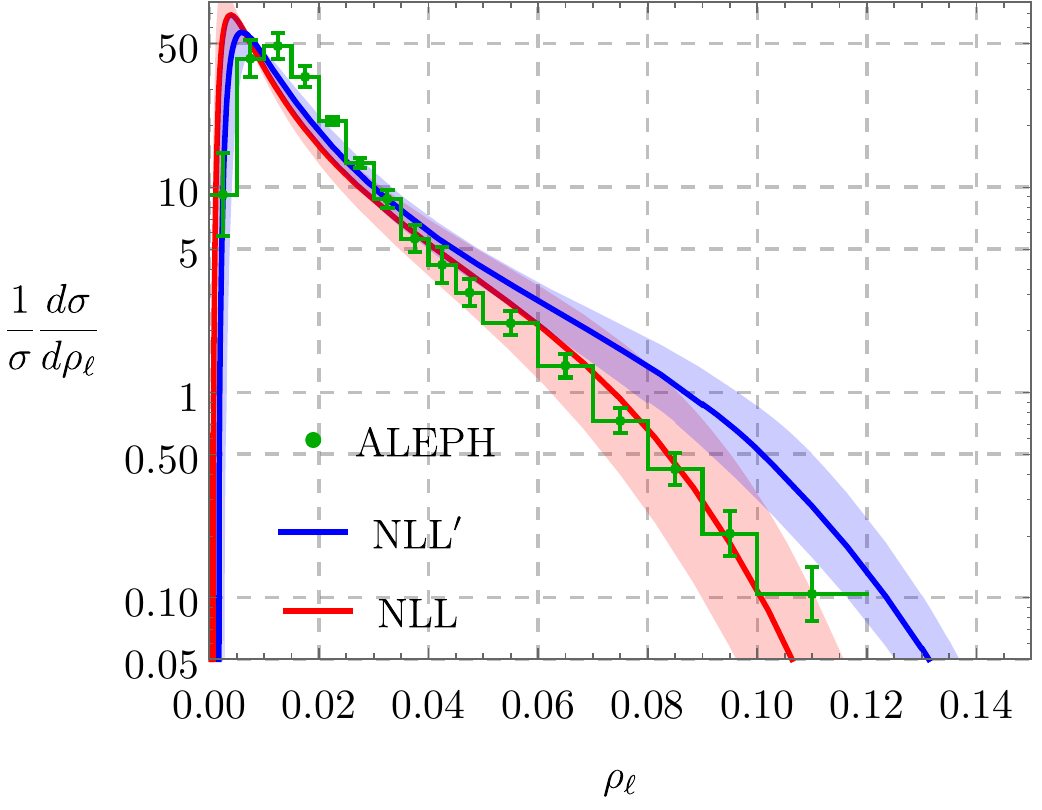} 
\end{tabular}
\caption{Light-jet and heavy-jet mass distribution in comparison to {\sc ALEPH} data \cite{Buskulic:1992hq}.\label{fig:LHJet_NLLp}}
\end{figure}

For completeness, we show in Figure~\ref{fig:LHJet_NLLp} numerical results for the heavy-jet mass $\rho_h$ and the light-jet mass $\rho_\ell$. The heavy-jet mass is global and provides a reference variable at the same accuracy, but free from all the complications which arise for the jet mass. From the difference of the heavy-jet mass and the jet mass we obtain the light-jet mass. This is more sensitive to the non-global structure and also only has a nontrivial distribution at $\mathcal{O}(\alpha_s^2)$ so that there is no matching at the accuracy we work. The end-point for the NLO light-jet mass is at $\rho_{\rm max} = 1/6$, which is achieved when the four parton momenta form a tetrahedron, and we use this as the endpoint in our profile function \eqref{profile}. From the plot, one observes that also the heavy-jet distribution is affected by nonperturbative effects in the peak region, however, the peak is at a larger $\rho$ value than for the jet mass itself. Not surprisingly, the worst description of the data arises for the light-jet mass distribution. At larger $\rho$ values the description is worse because the fixed-order result starts at $\mathcal{O}(\alpha_s^2)$ so that the matching corrections are beyond the accuracy of our computation. The peak region is not well described because it is in the nonperturbative regime and very narrow.

\section{Conclusion and Outlook\label{sec:concl}}

In this paper we analyzed non-global observables and, for the first time, went beyond a resummation of only the leading non-global logarithms. Specifically, we analyzed the single-logarithmic interjet energy flow at ${\rm LL}'$ and the double-logarithmic jet mass at ${\rm NLL}'$.
The prime indicates that we included the full next-to-leading-order corrections to the hard and soft functions, as well as the jet function in the case of jet mass. The practical implementation of these corrections is the main result of the present paper.  To achieve full NLL resummation for the interjet energy flow, and NNLL accuracy for the jet mass, we will need to also include the two-loop corrections the RG running, but we observe that the inclusion of the one-loop matching corrections already leads to an improved description of these observables. Since the jet mass peaks at a low value corresponding to a soft scale of $M_J^2/Q \approx 0.5\,{\rm GeV}$ for {\sc LEP} energies, the peak region is strongly affected by non-perturbative effects, similar to what is observed for other event shapes.

Due to the intricate structure of the soft emissions, factorization theorems for non-global observables and the associated RG evolution are much more complicated than in the global case. Instead of analytical computations, one needs to resort to a numerical Monte Carlo framework to perform the resummation. While the global heavy-jet mass only involves a soft function with two Wilson lines, the shower evolution for jet mass produces additional legs, and for low jet masses we can end up with soft emissions from hundreds of hard partons. 
%For example, for a low scale of $\mu_s=0.7\,{GeV}$  $t=0.1$ ($t=0.2$) and a collinear cutoff $\eta_{\cut} = 5$, we end up with with about 
However, concerning the NLO soft function, this is a minor complication, since we only connect pairs of legs at this accuracy.
Indeed, the inclusion of the NLO corrections to the soft function is a minor modification of the leading-logarithmic shower framework. Using the shower emissions  which end up in the veto region, we are able to compute the next-to-leading-order correction to the soft function in a general way, with almost no additional computer time.

The more involved part is the implementation of the NLO hard functions. These are in essence the usual real and virtual fixed-order corrections to the Born-level process, but individually suffer from collinear divergences. Computing them in dimensional regularization and renormalizing, one ends up with distributions in the angles of the hard partons which must be implemented into the Monte Carlo framework. We do this with a simple slicing scheme, which works well for two-jet production in $e^+ e^-$ but is certainly not the most efficient method. The problem of combining a parton shower with fixed-order results arises of course also for general purpose showers and elegant solutions such as {\sc MC@NLO} \cite{Frixione:2002ik} and {\sc POHWEG} \cite{Nason:2004rx} are available and have by now been fully automated. A complication in our case is that our shower systematically neglects small soft momenta and therefore does not conserve momentum. As a result, its kinematics is different from the one in the hard functions. While more work is needed on the NLO hard functions, let us note that we have achieved full automation for the leading-order hard functions in our previous paper \cite{Balsiger:2018ezi} by working with Les Houches event files generated by the tree-level generator in {\sc MadGraph5$\_$aMC@NLO}. The same code also provides NLO shower matching and it would be very interesting to adapt it to our shower.

An important next step is of course the inclusion of second-order corrections into the RG-running to achieve the full resummation of subleading non-global logarithms. The corresponding anomalous dimension matrix involves three types of corrections: Double real emissions, real-virtual terms and fully virtual two-loop corrections.  The relevant anomalous dimension matrix has been presented in a related framework by Caron-Huot \cite{Caron-Huot:2015bja}. We are working on determining the anomalous dimension also in our formalism. The implementation into a Monte Carlo framework will be nontrivial, because one needs to numerically handle the collinear singularities of the individual entries. There are a number of recent papers addressing the issue of double emissions in general parton showers \cite{Li:2016yez,Hoche:2017iem,Hoche:2017hno,Dulat:2018vuy}.

A second interesting challenge is the inclusion of finite-$N_c$ effects, especially for non-global observables at hadron colliders. Our RG-evolution framework is in the general class of showers characterized in \cite{Nagy:2007ty} and valid at finite $N_c$, but implementing the interference effects and complex phases which arise beyond $N_c \to \infty$ is challenging. Interesting progress towards the computation of such corrections has been made in \cite{Platzer:2012np,Martinez:2018ffw}.

We have analyzed two simple non-global observables in the present paper. This is a first step, but our ultimate goal is of course to use the same methods to understand jet structure at the LHC. For narrow jets, the non-global structure actually factorizes into a structure for each separate jet \cite{Dasgupta:2012hg,Becher:2015hka,Becher:2016mmh}. Boosting our hemisphere jet mass result such that the left hemisphere transforms into a cone of radius $R$, one immediately obtains the non-global structure of the jet mass for an LHC jet of this radius. It will be interesting to analyze such observables in the future.

\begin{acknowledgments}	
The research of T.B.\ is supported by the Swiss National Science Foundation (SNF) under grant 200020\_182038. T.B.\ and D.Y.S.\ would like to express a special thanks to the Mainz Institute for Theoretical Physics (MITP) for its hospitality and support. D.Y.S.\ acknowledges helpful discussions with Pier Monni on numerical Monte Carlo resummation methods. The authors thank Rudi Rahn and Thomas Rauh for comments on the manuscript. 

\end{acknowledgments}

\appendix

\section{Hard and soft functions for interjet energy flow}\label{IJhs}

\subsection{Hard functions in the slicing scheme}

\begin{figure}[t]
 \includegraphics[width=0.45\textwidth]{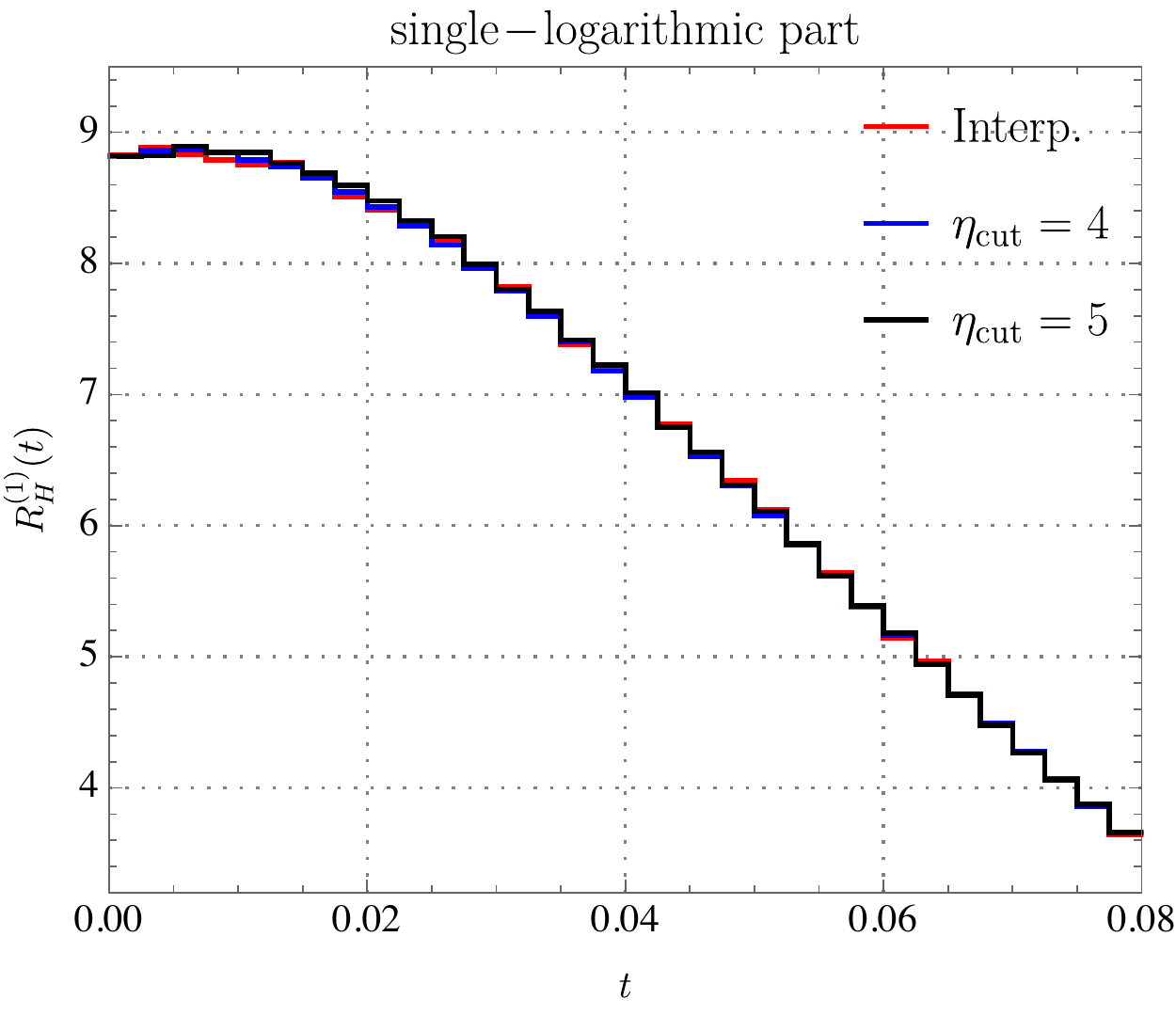}~~~~~
 \includegraphics[width=0.45\textwidth]{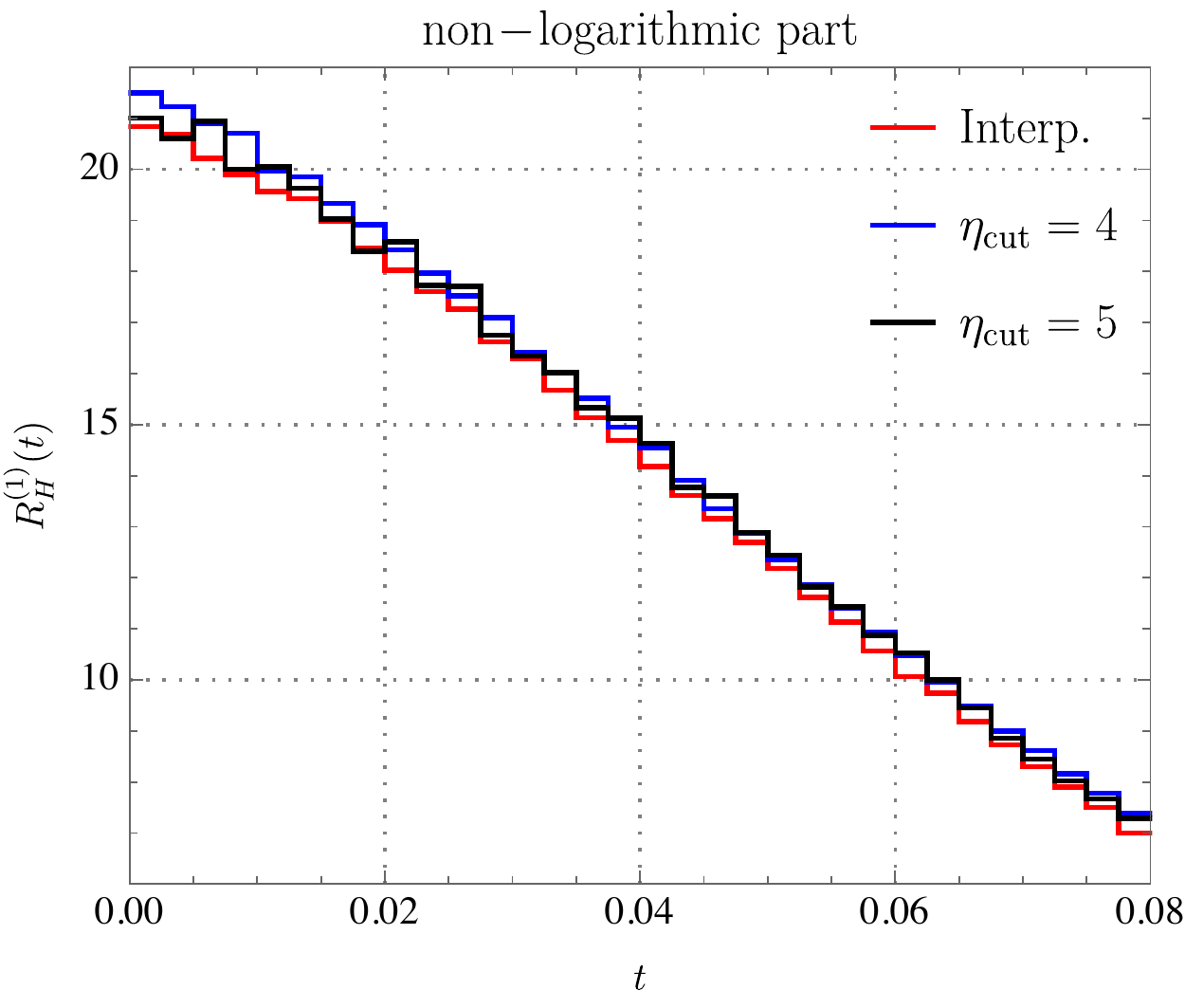}
 \caption{Numerical comparison among different Monte Carlo implementations of the one-loop hard corrections $R_H^{(1)}$ to the gap fraction. The red line corresponds to the interpolation method, the other two are obtained using the slicing method with different values of the cutoff $\eta_{\rm cut}$. Left:  Coefficient of the single logarithmic part. Right: Non-logarithmic terms.  \label{fig:cutoff_dep}}
 \end{figure}

We discussed in Section~\ref{sec:interjet} that one can use a simple slicing method to implement the plus distribution terms inside hard function $\mathcal{H}_3^{(1)}(u,v)$ into the shower. In the main text, we have explained the procedure using the toy example \eqref{slicing} in which we disregarded the $u$ dependence. In this appendix we now provide the full expression for the hard function. As explained in Section~\ref{sec:interjet}, we can directly integrate over $u$ for the $\delta(v)$ terms and combine them with $\widehat{\mathcal{S}}_2$. Since there are no singularities inside Region~III, we only give expressions for Regions I and II,
\begin{align}\label{H3Is}
\mathcal{H}_{3, {\rm I}}^{(1)}(u,v, Q,\mu) =&~ C_F \Bigg\{ \Bigg[ 4\ln^2\frac{\mu}{Q} + \ln\frac{\mu}{Q}\left( \frac{7}{2} + 8 \ln 2 - 8\ln v_0 \right) + 4 - \frac{\pi^2}{6} + \frac{7}{2}\ln2 + 4\ln^2 2  \nno \\
& \hspace{-2cm}+ \ln v_0 \left[ \frac{ - 7 + 2u_0 + 5u_0^2}{2(1+u_0)^2} - 8\ln 2 + 8 \ln(1+u_0) \right] + 4\ln^2 v_0 \Bigg]\delta(u)\delta(v)  \nno \\
& \hspace{-2cm} + F(0,v) \delta(u)  \frac{\theta(v - v_0)}{v} \left[\ln u_0 -  \ln \frac{\mu}{Q}   +  \ln v \right]  + F(u,v) \frac{\theta(u - u_0)\theta(v - v_0)}{uv }\Bigg\}\Theta_{\rm in}(v)\,, \\
\mathcal{H}_{3, {\rm II}}^{(1)}(u,v, Q,\mu) =&~ C_F \Bigg\{ \Bigg[ \left(\ln\frac{\mu}{Q} -\ln v_0\right)\left( \frac{5}{2} - 8\ln2 \right) + 3 - \frac{2\pi^2}{3} + \frac{5}{2}\ln2 - 4\ln^2 2 \Bigg]\delta(u)\delta(v)  \nno \\
& \hspace{-2cm}  + G(u,v) \frac{\theta(v - v_0)}{v }\Bigg\} \Theta_{\rm in}(v)\,, 
\end{align}
where the cutoffs on $u$ and $v$ are chosen as $u_0=v_0= e^{-\eta_{\rm cut} -1}$ in the parton shower code. The cutoff $\eta_{\rm cut}$ is imposed in the parton shower on the rapidity of the emitted hard partons. It can be imposed in the lab frame or in the COM frame of the emitting dipole, see \cite{Balsiger:2018ezi} for more discussions. We have checked that the cutoff dependence can be neglected, as can be observed in Figure~\ref{fig:cutoff_dep} in which we show a numerical comparison between the results based on interpolating the soft function and the slicing method for different cutoffs. 

 \subsection{One-loop soft functions}

At the one-loop level, virtual corrections from soft gluons are scaleless (and therefore vanish in dimensional regularization), and we only need to include real-emission contributions. The soft function consists of a $d$-dimensional integral with phase-space cuts which ensure that the real emission is outside the jets (the inside part is again scaleless). The relevant soft integral is given by
\begin{align}
	\frac{\alpha_s}{4\pi} &\bm{S}^{(1)}_m(\{\underline{n}\},Q_0,\e)=\nno \\
	& - g_s^2 \sum_{(ij)}\,\,\bm{T}_{i,L}\cdot\bm{T}_{j,R}  \, \tilde{\mu} ^{4-d}\int\frac{d^d k}{(2\pi)^{d-1}}\frac{n_i \cdot n_j}{n_i \cdot k \, n_j \cdot k}\delta(k^2)\theta(k^0)\theta(Q_0-v\cdot k)\Theta_{\rm out}(n_k),
\end{align}
with $\tilde{\mu} = e^{\gamma_E}\mu ^2/(4\pi)$ with $v^\mu = (1,0,0,0)$ and $v\cdot k = k^0$. To evaluate the contribution of the $(n_i,n_j)$ dipole, we Lorentz transform into a frame where the vectors $n_i$ and $n_j$ are back-to-back and the reference vectors take the form 
\begin{align}
\hat n_i^\mu &= \frac{M}{2}(1,0,0,1)\, , & \hat n_j^\mu &=  \frac{M}{2}(1,0,0,-1)\,,  &  \hat v^\mu  &=  \frac{2}{M}(1,0,\beta,0)\,,
\end{align}
where $M^2 = 2\, n_i \cdot n_j$ is the invariant mass of the dipole pair, and $\beta=\sqrt{1-M^2/4}$.  In this frame, we parameterize the integration momentum as
\begin{align}
k = \hat k_T ( \cosh \hat y, \sin \hat \phi, \cos \hat \phi, \sinh \hat y ). 
\end{align} 
With $d=4-2\e$, the integral then reads
\begin{align}
\int \! d^d k &\frac{\hat n_i \cdot \hat n_j}{\hat n_i \cdot k \, \hat n_j \cdot k}\delta(k^2)\theta(k^0)\theta(Q_0-k \cdot \hat v )\Theta_{\rm out}(\hat n_k) = \nno \\
& \frac{\Omega_{d-3}}{2}\, \int_0^\infty \frac{d \hat k_T}{\hat k_T^{1+2\e}}  \int_{-\infty}^{\infty} \! d\hat{y} \int_0^{\pi} \! d\hat{\phi}\, |\sin\hat{\phi} |^{-2\e} 
\theta(Q_0-k \cdot \hat{v})\Theta_{\rm out}(\hat{n}_k)\,,
\end{align}
where $\Omega_{d}$ is the surface of the $d$-dimensional unit sphere and $\Omega_1 = 2$. Introducing the auxiliary function $f_{ij}$ via
\begin{equation}
 k \cdot \hat{v} = f_{ij}(\hat{y},\hat{\phi})\, \hat k_T = \frac{2}{M} \big(\cosh \hat y-\beta\cos\hat\phi \big) \hat k_T \, ,
 \end{equation}
we can perform the integral over $\hat k_T$. This integration yields a soft divergence, which is renormalized away in the  ${\rm \overline{MS}}$ scheme. After expanding in $\e$ we then immediately arrive at expression \eqref{eq:SmNLOb} which only involves a finite angular integration which we perform with the parton shower, which generates its emissions using the variables $\hat{y}$ and $\hat{\phi}$. 

\section{Monte Carlo algorithm for the interjet energy flow}\label{IJmc}

The inclusion of the NLO soft function is only a minor modification of the algorithm for LL resummation. In fact, the first three steps are identical to what was shown in Appendix B of \cite{Balsiger:2018ezi}.  The only difference arises in the last step, where we also compute the soft function. To record the results of the shower, we fill three histograms: $h_U$ contains the LL evolution, $h_L$ the coefficient of the logarithm of the soft function \eqref{eq:SmNLOb} and $h_c$ its non-logarithmic part.
 
The shower algorithm for the evolution of the function $\bm{\mathcal{H}}_2^{(0)}(\{n_1,n_2 \},Q,\mu_h)$ to lower scales involves the following steps:
\begin{enumerate}
\item  \label{one}Start at evolution time $t=0$ from an initial event $E$ with vectors $\{  n_1,  n_2 \}$ and weight $w=1$.
\item  \label{two} Generate a random time step $\Delta t$ according to the probability distribution $\mathcal{P}_E(t)= V_E \exp(-V_E \Delta t)$, and insert the event weight $w$ into the histogram $h_U$ at time $t+\Delta t$.
\item \label{three} Choose a dipole associated with a pair of neighbouring vectors $n_i$ and $n_j$ in $E$ with probability $V_{ij}/V_E$. Generate a new random vector $n_k$ and multiply the weight by the factor $R_{ij}^k/V_{ij}$, expressed in the random variables chosen to generate the direction of the new vector $n_k$. 
\item \label{four} If $n_k$ is outside the veto region, add this new vector to the event which then becomes $E'=\{n_1,\cdots,n_{i},n_k,n_j,\cdots,n_2\}$, multiply the weight by a factor $V_E/V_{E'}$ and return to Step~\ref{two}. Otherwise, add the weight factors
\begin{align}
w  ~~~~ {\rm and} ~~~~ \ln\frac{2\, |\sin\hat\phi|}{f_{ij}(\hat\phi,\hat y)} w
\end{align}
to $h_L$ and $h_c$ at time $t$, go to Step~\ref{one} and start a new event.
\end{enumerate}
In terms of these histograms, the soft function correction reads
\begin{equation}
\frac{\alpha_s (\mu_s)}{4\pi}\sum_{m=2}^\infty \big\langle  \bm{\mathcal{H}}_2^{(0)} \, \otimes \,  \bm{U}_{2m} \,\hat{\otimes}\, \bm{\mathcal{S}}_m^{(1)} \big\rangle =  \frac{N_c}{2} \frac{\alpha_s (\mu_s)}{4\pi} \left[ -4 \ln\frac{\mu}{Q_0} h_L(t) + 4\, h_c(t) \right],
\end{equation}
while the LL evolution factor is
\begin{equation}
\sum_{m=2}^\infty \big\langle  \bm{\mathcal{H}}_2^{(0)} \, \otimes \,  \bm{U}_{2m} \,\hat{\otimes}\, \bm{1} \big\rangle =   h_U(t) \,.
\end{equation}
We discussed the implementation of $\bm{\mathcal{H}}_3^{(1)}$ in the main text. The shower algorithm is the same as the one described above, up to the fact that one starts the shower with a three parton configuration and does not need to compute the one-loop soft function.

\section{Hard and soft functions for the jet mass} \label{LjetHS}

In this appendix we list one-loop ingredients for the jet mass.  The ingredients are closely related to the ones relevant for the interjet energy flow, but the notation is somewhat different. For the jet mass, the hard function $\mathcal{H}_m^{q(1)}$ denotes the configuration with a quark on the left and $m$ partons in the right hemisphere, while $m$ simply counts the total number of hard partons for the interjet energy flow. 
In the large $N_c$ limit the renormalized one-loop hard function $\mathcal{H}_1^i$ are thus given by
\begin{align}\label{eq:H1jet}
\mathcal{H}_1^{q(1)}(\hat \theta_1,Q,\mu)= \mathcal{H}_1^{\bar q(1)}(\hat \theta_1,Q,\mu)= \frac{1}{2} \,\delta(\hat\theta_1) H_2(Q^2,\mu)\,.
\end{align} 
The factor of one half is present because the LO total cross section is a sum of two identical contributions with the quark and anti-quark in the left hemisphere, respectively. The $\delta$-function of $\hat \theta_1 = \tan(\theta_1/2)$ with $\theta_i$ ensures that the right parton flies along $n$-direction, opposite to the left parton along $\bar{n}$.

Since the thrust axis points along the opposite direction of the most energetic parton for a three-jet configuration, also the hard functions $\mathcal{H}_1^{i(1)}(\hat \theta_1,Q,\mu)$ are the same as for the interjet energy flow. We use the same variables $u$ and $v$ introduced for the interjet energy flow to parameterize the angular variables in order to resolve the overlapping divergences inside the angular integration:
\begin{align}
{\rm Region~I}\,( \theta_g > \theta_{\bar q}) :~~& v= \tan\frac{\theta_g}{2}, ~~~ u \,  v = \tan\frac{\theta_{\bar q}}{2}, \nno \\
{\rm Region~II}\,(\theta_g < \theta_{\bar q}):~~& v= \tan\frac{\theta_{\bar q}}{2}, ~~~ u \,  v = \tan\frac{\theta_{g}}{2}, \nno\\
{\rm Region~III}\,( \theta_q > \theta_{\bar q}) :~~& v= \tan\frac{\theta_q}{2}, ~~~ u \,  v = \tan\frac{\theta_{\bar q}}{2} ,  \nno
\end{align}
where the regions are depicted in Figure~\ref{fig:H3_dia}. For the jet mass case, we no longer impose a cone constraint (i.e. we can set $\delta=1$), but we need to add the constraint
 \begin{equation}
\Theta_T(u,v) = \theta\!\left[\sqrt{1+u^2v^2} - (1+u)\,v\right]
\end{equation} 
to ensure that the thrust axis does not flip. On the level of the bare function, this constraint was given in (4.4) of \cite{Becher:2016omr}, but was trivially fulfilled for our choice of the cone angle. Due to this constraint, the angle of any parton to the thrust axis cannot be larger than $\frac{\pi}{3}$.

Performing the variable transformation and writing the angular convolution as integrals over $u$ and $v$ as in \eqref{H3S3} we have
\begin{align}
\mathcal{H}_{2, {\rm I}}^{q(1)}(u,v,Q,\mu) =&~ \frac{1}{2} \mathcal{H}_{3, {\rm I}}^{(1)}(u,v, Q,\mu) \, \Theta_T(u,v)  \, ,\\
\mathcal{H}_{2, {\rm II}}^{q(1)}(u,v,Q,\mu) =&~ \frac{1}{2} \mathcal{H}_{3,{\rm II}}^{(1)} (u,v,Q,\mu)\, \Theta_T(u,v) \,, \\
\mathcal{H}_{2, {\rm III}}^{g(1)}(u,v,Q,\mu) =&~ \frac{1}{2} \mathcal{H}_{3,{\rm III}}^{(1)}(u,v,Q,\mu)\, \Theta_T(u,v) \,,
\end{align}
where the factor $\frac{1}{2}$ has the same source as in \eqref{eq:H1jet} and the interjet functions were given in \eqref{H3I}, \eqref{H3II} and \eqref{H3III}. The anti-quark hard function $\mathcal{H}_2^{\bar q(1)}$ is equal to the quark function. For the gluon function, there is also a region  $\theta_{\bar q} > \theta_q $ which is parameterized analogously and gives an identical contribution.

As explained in \cite{Becher:2016omr}, the soft function for the light-jet mass is directly related to the coft function in Sterman-Weinberg dijet cross section defined in \cite{Becher:2016mmh}. In Laplace space, we have
\begin{align}\label{eq:SmNLOJet}
   \frac{\alpha_s}{4\pi}\widetilde{\bm{S}}_m^{(1)}(\{\underline{n}\},\tau,\e)& =   \nno \\
   & \hspace{-2.8cm}
   - g_s^2\,\tilde{\mu}^{2\e} \sum_{(ij)}\,\bm{T}_{i,L}\cdot\bm{T}_{j,R} 
  \int\frac{d^{d}k}{(2\pi)^{d-1} }\delta(k^2)\theta(k^0) e^{-\bar n \cdot k/(\tau e^{\gamma_E})}\,\frac{n_i\cdot n_j}{n_i\cdot k\,n_j\cdot k}\theta(n\cdot k-\bar{n}\cdot k)\, .
\end{align}
The evaluation of this expression proceeds along the same lines as for the interjet energy flow case derived in detail in Appendix~\ref{IJhs}. If both emitting partons are in the right hemisphere, the renormalized one-loop result is given by
\begin{align}\label{eq:SmJet}
\widetilde{\bm{S}}_m^{(1)}(\{\underline{n}\},\tau,\mu) = \frac{N_c}{2} \sum_{i,j=1}^m \delta_{i,j\pm 1} \int d\hat y \, \frac{d\hat\phi}{2\pi} \left[ -4 \ln\frac{\mu}{\tau} + 4\ln\frac{2\, |\sin\hat\phi|}{g_{ij}(\hat\phi,\hat y)} \right] \Theta_{\rm L}^{\rm lab}(\hat y,\hat\phi) \, , 
\end{align}
with the measurement function $\Theta_{\rm L}^{\rm lab}(\hat y,\hat\phi)$ constraining the soft radiation to the left hemisphere, and a function 
\begin{align}\label{eq:gfun}
g_{ij}(\hat\phi,\hat y) =&\, \frac{1}{\beta  M} \Big[2\beta  \cosh \hat y+\beta  e^{\hat y} \tanh y_i+\beta  e^{-\hat y}\tanh y_j-\cos \hat \phi  \left[2 \beta ^2+\tanh y_i+\tanh y_j\right] \nno \\
& +\text{sech}  y_i \,\text{sech} y_j \sin \hat \phi  \sin (\phi_i -\phi_j) \Big]. 
\end{align}
If one of the two partons is on the left, the function has a collinear divergence, which can be subtracted, as detailed in Section~\ref{sec:jetmass}. The subtraction was given in \eqref{eq:globalSubtr}.

\section{Monte Carlo algorithm for the jet mass distribution}\label{LJmc}

In this appendix we provide the Monte Carlo algorithm used for jet mass resummation, which is also applicable for other non-global observables with soft-collinear double logarithms. Compared to interjet energy flow, we need to subtract the global anomalous dimension and the one-loop global soft function. As for the interjet energy case, we fill three histograms: $h_U$ contains the LL evolution, $h_L$ the coefficient of the logarithm of the soft function \eqref{eq:SmJet} and $h_c$ its non-logarithmic part. 

The algorithm for evolving $\mathcal{H}_{1}^{q}$ to lower scales  involves the following steps:
\begin{enumerate}
\item  \label{oneJet} Start at evolution time $t=0$ from an initial event $E$ with vectors $\{  \bar n,  n_1 \}$ and weight $w=1$.
\item  \label{twoJet} Generate a random time step $\Delta t$ according to the probability distribution $\mathcal{P}_E(t)= \overline{V}_{\! E} \exp(-\overline{V}_{\! E}\, \Delta t)$, and insert the event weight $w$ into the histogram $h_U$ at time $t+\Delta t$.
\item \label{threeJet} Choose a dipole associated with a pair of neighbouring vectors $n_i$ and $n_j$ in $E$ with probability $V_{ij}/ V_E$. Generate a new random vector $n_k$ and multiply the weight by the factor $R_{ij}^k/V_{ij}$, expressed in the random variables chosen to generate the direction of the new vector $n_k$. 
\item \label{fourJet} If $n_k$ is in the right hemisphere, add this new vector to the event so that $E'=\{\bar n,\cdots,n_{i},n_k,n_j,\cdots,n_1\}$, multiply the weight by a factor $ V_E/\overline V_{E'}$ and return to Step~\ref{twoJet}. If $n_k$ is in the left hemisphere and was emitted from dipole $(\bar n,n_j)$, we need to  subtract the global one-loop soft function $\widetilde S^{q\,(1)}_G$ in equation \eqref{eq:globalSubtr}. This is achieved with the weight factors
\begin{align}
\left[1 - X(\hat\phi,\hat y) \right] w  ~~~~ {\rm and} ~~~~ \ln\frac{2\, |\sin\hat\phi|}{g_{0j} (\hat\phi,\hat y)} \left[1 - X(\hat\phi,\hat y) \right] w\,,
\end{align}
which are added to the histograms $h_L$ and $h_C$ at time $t$. After filling the histograms go to Step~\ref{oneJet} and start a new event.  Otherwise, add the unsubtracted weight factor 
\begin{align}
w  ~~~~ {\rm and} ~~~~ \ln\frac{2\, |\sin\hat\phi|}{g_{ij}(\hat\phi,\hat y)} w
\end{align}
to the respective histograms, go to Step~\ref{oneJet} and start a new event. 
\end{enumerate}
The quantity $\overline {V}_{\! E}$ denotes the subtracted global anomalous dimension $\overline V_{\! E} = V_{ E} - V_0$, where $V_0$ is the large-$N_c$ result for the subtraction \eqref{subtraction} obtained by replacing the Casmir operator $C_i$ in this equation by $N_c/2$ for a quark jet, or $N_c$ for a gluon jet, respectively.

\section{Ingredients for jet mass resummation}\label{adapp}
For convenience, we collect here the perturbative results for ingredients used in the resummation formula for jet mass distribution. The evolution factors at NLL accuracy are given by
\begin{align} 
S \left( \nu  , \mu \right) & = \frac { \gamma _ { 0 }^{\rm cusp} } { 4 \beta _ { 0 } ^ { 2 } } \left\{ \frac { 4 \pi } { \alpha _ { s } \left( \nu  \right) } \left( 1 - \frac { 1 } { r } - \ln r \right) + \left( \frac { \gamma _ { 1 }^{\rm cusp} } { \gamma _ { 0 } ^{\rm cusp} } - \frac { \beta _ { 1 } } { \beta _ { 0 } } \right) ( 1 - r + \ln r ) + \frac { \beta _ { 1 } } { 2 \beta _ { 0 } } \ln ^ { 2 } r \right\} \, , \nno\\
 A _ { \gamma } \left( \nu , \mu \right) & = \frac { \gamma _ { 0 } } { 2 \beta _ { 0 } } \ln r \, ,\end{align}
with $r = \alpha _ { s } ( \mu ) / \alpha _ { s } \left( \nu \right)$. The expressions of the anomalous dimensions used in our paper are
\begin{align}
& \gamma _ { 0 }^{\rm cusp} = 4 \, , \quad \gamma _ { 1 }^{\rm cusp} = \left( \frac { 268 } { 9 } - \frac { 4 \pi ^ { 2 } } { 3 } \right)  C _ { A } - \frac { 80 } { 9 } T _ { F } n _ { f }\,, \nno \\
& \gamma _ { 0 } ^ { J_q } = - 3 C _ { F }\, ,\quad c_1^{J_q} = C_F \left(7 - \frac{2\pi^2}{3}\right), \quad \gamma_{0}^{J_g} = - \beta_0\, , \nno \\
&  \beta _ { 0 } = \frac { 11 } { 3 } C _ { A } - \frac { 4 } { 3 } T _ { F } n _ { f }\, , \quad \beta _ { 1 } = \frac { 34 } { 3 } C _ { A } ^ { 2 } - \frac { 20 } { 3 } C _ { A } T _ { F } n _ { f } - 4 C _ { F } T _ { F } n _ { f }\, . 
\end{align}
The LO integrated jet mass distribution is written as
\begin{align}\label{eq:LOjetmass}
\frac{1}{\sigma_0}  \int_0^{\rho} \!d \bar \rho\, \frac{d\sigma^{\rm LO}}{d \bar \rho} = & \, 1+  C_F  \frac{\alpha_s}{2\pi}  \bigg[ -\ln^2\rho - \frac{3}{2} \ln\rho + \frac{1}{4} + \frac{\pi^2}{6} - 2 \, {\rm Li}_2\!\left(\frac{\rho }{1-\rho }\right)+\frac{9 \rho ^2}{4} + 3 \rho  \nno \\
&  -  \ln^2(1-\rho ) + \frac{3}{2} (1-2 \rho ) \ln (1-2 \rho )  +\Big[3 \rho + 2 \ln (1-\rho )\Big] \ln \rho \bigg]  .
\end{align}
The integrated light-jet mass distribution is trivial at this order
\begin{align}\label{eq:LOlightjetmass}
 \int_0^{\rho} \! d \bar \rho\, \frac{d\sigma^{\rm LO}}{d \bar \rho} =\sigma_0 \left( 1+  C_F  \frac{3\alpha_s}{4\pi} \right)= \sigma \,,
\end{align}
because the light jet has zero mass for three partons.

\end{document}